\documentclass[times,sort&compress,3p]{elsarticle}
\journal{Journal of Multivariate Analysis}
\usepackage[labelfont=bf]{caption}

\usepackage{amsmath,amsfonts,amssymb,amsthm,booktabs,color,epsfig,graphicx,hyperref,url,setspace,enumerate,mathrsfs}
\usepackage{natbib,epsfig,graphicx,pdfpages}
\usepackage{rotating}
\usepackage{float}
\usepackage{bm}
\usepackage{ulem}
\usepackage{CJK}
\usepackage{natbib}
\usepackage{algorithm}
\usepackage{algorithmic}
\usepackage{booktabs}
\usepackage{bm}
\usepackage{subfigure}
\usepackage{makecell}
\usepackage{hyperref}
\usepackage{multirow}
\usepackage{threeparttable}
\usepackage{xr}
\usepackage{mathtools}

\def\beq{\begin{equation}}
	\def\eeq{\end{equation}}
\def\beqr{\begin{eqnarray}}
	\def\eeqr{\end{eqnarray}}
\def\beqrs{\begin{eqnarray*}}
	\def\eeqrs{\end{eqnarray*}}
\def\bet{\begin{theorem}}
	\def\eet{\end{theorem}}
\def\bel{\begin{lemma}}
	\def\eel{\end{lemma}}
\def\bes{\begin{step}}
	\def\ees{\end{step}}
\def\bep{\begin{proposition}}
	\def\eep{\end{proposition}}
\def\bg{\begin{figure}[tbph]\begin{center}}
		\def\eg{\end{center}\end{figure}}

\def\bc{\begin{center}}
	\def\ec{\end{center}}

\def\wt{\widetilde}

\def\wh{\widehat}

\def\mD{\mathcal D}
\def\mQ{\mathcal Q}
\def\mR{\mathbb{R}}
\def\mS{\mathcal S}

\def\mL{\mathcal L}
\def\mB{\mathcal B}

\def\1{\mbox{\boldmath $1$}}

\def\mO{\mathcal O}

\def\mT{\mathcal T}

\def\var{\mbox{var}}

\def\cov{\mbox{cov}}
\def\tr{\mbox{tr}}

\def\argmax{\mbox{argmax}}

\def\diag{\mbox{diag}}

\newcommand{\Rmnum}[1]{\expandafter\@slowromancap\romannumeral #1@}
\makeatother

\theoremstyle{plain}
\newtheorem{theorem}{Theorem}

\newtheorem{proposition}{Proposition}
\newtheorem{lemma}{Lemma}

\theoremstyle{definition}

\begin{document}

\begin{frontmatter}

\title{A Latent Factor Model for High-Dimensional Binary Data}

\author[1]{Jiaxin Shi}
\author[1]{Yuan Gao\corref{mycorrespondingauthor}}
\author[2]{Rui Pan}
\author[1]{Hansheng Wang}

\address[1]{Guanghua School of Management, Peking University, Beijing, China}
\address[2]{School of Statistics and Mathematics, Central University of Finance and Economics, Beijing, China}

\cortext[mycorrespondingauthor]{Corresponding author. Email address: \url{ygao_stat@outlook.com}}

\begin{abstract}
	
In this study, we develop a latent factor model for analysing high-dimensional binary data. Specifically, a standard probit model is used to describe the regression relationship between the observed binary data and the continuous latent variables. Our method assumes that the dependency structure of the observed binary data can be fully captured by the continuous latent factors. To estimate the model, a moment-based estimation method is developed. The proposed method is able to deal with both discontinuity and high dimensionality. Most importantly, the asymptotic properties of the resulting estimators are rigorously established. Extensive simulation studies are presented to demonstrate the proposed methodology. A real dataset about product descriptions is analysed for illustration.
\end{abstract}


\end{frontmatter}

\section{Introduction\label{sec:1}}

Factor modelling is a powerful tool for analysing high-dimensional data \citep{sammon1969nonlinear,meng1996fitting,bartholomew2011latent}. The key idea of factor modelling is to compress high-dimensional data into low-dimensional features. In the meanwhile, the relevant information should be preserved as much as possible. Due to its theoretical and practical importance, much research effort has been devoted in this regard. For instance, \cite{1956statistical} discussed likelihood-based statistical inference for factor analysis. It was further extended by \cite{1988asymptotic} under general conditions. More recently, factor models were used by \cite{fan2008,fan2011high} for high-dimensional covariance matrix estimation, while \cite{wang2012} employed them for ultrahigh dimensional variable screening. Dynamic factor models were studied by \cite{1985dynamic} and \cite{2018dynamic} for multivariate time series analysis. It was further generalized by \cite{2021dynamic} in state space models. Despite the practical usefulness and theoretical significance of those factor models, they all suffer from one common limitation. That is, they are almost developed for continuous data. This limitation leaves the problem of factor modelling for high-dimensional binary data open for discussion.

In fact, high-dimensional binary data are encountered frequently in a wide range of research areas. Those important areas include but are not limited to text mining \citep{zhao2005hierarchical}, biology \citep{gui2005penalized}, medical science \citep{fan2011application}, social science \citep{gollini2014mixture}, and many others. However, such data are hard to handle via classical methods due to their discontinuity and high dimensionality \citep{cox1972analysis,bartholomew2011latent}. To solve this problem, \cite{bartholomew1980factor} introduced latent factor models for ordered categorical data though the conditional independence of the latent continuous variables. A Poisson factor model for multivariate count data was further developed by \cite{wedel2003factor}. The problem of dimension reduction for high-dimensional binary data was investigated by \cite{schein2003generalized}, \cite{lee2010sparse}, \cite{davenport20141}, and \cite{2020dimensionality}, which include binary PCA, logistic PCA, 1-bit low rank matrix completion, and others. Although these pioneering methods are practically useful, their statistical properties remain largely unknown. Key statistical properties, such as consistency, have yet to be established. Therefore, we are motivated to develop a principled factor modelling approach for high-dimensional binary data with solid theoretical support.

To this end, we propose a latent factor model for high-dimensional binary data. Our method comprises two important components. The first one is a standard probit model, which is used to describe the relationship between the observed binary data and the continuous latent variable \citep{bliss1935}. The second component is a high-dimensional latent factor model \citep{1978contributions}, which admits a continuous form. Thus, the dependency structure of the observed binary data can be fully captured by the continuous latent factors. This is not the first attempt to treat the factor analysis of binary data. However, the earlier attempts primarily focus on computational techniques rather than the fundamental statistical properties \citep{schein2003generalized,davenport20141,2020dimensionality}. In this work, a simple method of moment estimation is developed to estimate the model and its supporting asymptotic theory is rigorously established. 

To summarize, this study makes two important contributions to the literature. The first is the development of a probit-model-based factor model for high-dimensional binary data analysis. The second contribution is the introduction of a moment-based estimation method with rigorous theoretical support. To demonstrate the finite sample performance of our proposed method, extensive simulation studies are conducted. Furthermore, we also present a real data example involving a Chinese text dataset of 54,000 product descriptions,  each of which can be represented by a high-dimensional binary vector of dimension 200.

The rest of the article is organized as follows. Section 2 develops the latent factor model for high-dimensional binary data. The asymptotic theory is also established. The numerical studies are presented in Section 3, including the simulation experiments and a real data example of product description analysis. Finally, Section 4 concludes the article with a brief discussion. All technical proofs are contained in Appendix.

\section{Methodology\label{sec:2}}

\subsection{A Probit Model}

Let $ Y_i=(Y_{i1},\ldots,Y_{ip})^{\top}\in\mR^p $ be a $p$-dimensional binary random vector with $Y_{ij}\in\{0,1\}$ for every $1\leq i\leq n $ and $1\leq j\leq p $, where $n$ is the number of observations. For a given $i$, different $Y_{ij}s$ are mutually correlated. To statistically model this dependency structure, we assume that different $Y_{ij}s$ admit a standard probit model \citep{bliss1935,nelder1972,mccullagh2019} as
\begin{equation}\label{2.1}
	Y_{ij} = I\big(e_{ij}>c_j\big),
\end{equation}
where $ I(\cdot)$ is an indicator function and $e_{ij}$ is a standard normal variable. Moreover, $ c_j\in\mR^1 $ is an unknown parameter. Write $e_{i} = (e_{i1},\ldots,e_{ip})^{\top}\in\mR^p$. We then assume that $e_{i}$ is a multivariate normal random vector with $ E(e_{i})=0 $ and $ \cov(e_{i})=\Sigma=(\sigma_{j_1j_2})\in\mR^{p\times p} $. Then, we should have $ \sigma_{jj}=1 $ for $ 1\leq j\leq p $ and $|\sigma_{j_1j_2}|<1$ for $j_1\neq j_2$. Accordingly, the dependency structure of $Y_{ij}s$ is fully captured by $\Sigma$. We next consider how to estimate $\Sigma$. 

To estimate $\Sigma$, we should start with the estimation of $c_js$. Specifically, denote $P_j=P\big(Y_{ij}=1\big)= P\big(e_{ij}>c_j\big)=\Phi\big(-c_j\big)$ for every $ 1\leq j\leq p $, where $c_j=-\Phi^{-1}\big(P_{j}\big)$ and $\Phi(\cdot)$ stands for the cumulative distribution function of a standard normal distribution. Therefore, a simple moment estimator for $c_j$ can be constructed as $\wh{c}_j=-\Phi^{-1}\big(\wh{P}_j\big)$, where $\wh{P}_j=\sum_{i=1}^{n}I\big(Y_{ij}=1\big)/n$. To estimate $\sigma_{j_1j_2}$ with $j_1\neq j_2$, we next consider $P_{j_1j_2}=P\big(Y_{ij_1}=Y_{ij_2}=1\big)=P\big(e_{ij_1}>c_{j_1}, e_{ij_2}>c_{j_2}\big)$. Note that $ (e_{ij_1},e_{ij_2})$ is a bivariate normal random vector with $ E(e_{ij_1})=E(e_{ij_2})=0, \var(e_{ij_1})=\var(e_{ij_2})=1$, and $ \cov(e_{ij_1},e_{ij_2})=\sigma_{j_1j_2}$. Following \cite{kotz2004}, we denote $P_{j_1j_2}$ as $\ell\big(c_{j_1},c_{j_2};\sigma_{j_1j_2}\big)$. Then, we have  
\begin{equation}\label{2.2}
	\ell\big(c_{j_1},c_{j_2};\sigma_{j_1j_2}\big)=\dfrac{1}{2\pi\sqrt{1-\sigma_{j_1j_2}^2}}\int_{c_{j_1}}^{\infty}\int_{c_{j_2}}^{\infty}\exp\Bigg\{-\dfrac{x^2+y^2-2\sigma_{j_1j_2}xy}{2\big(1-\sigma_{j_1j_2}^2\big)}\Bigg\}dydx.
\end{equation}
Once $\wh{c}_{j_1}$ and $\wh{c}_{j_2}$ are obtained, a natural moment estimator for $\sigma_{j_1,j_2}$ can be defined as $\wh{\sigma}_{j_1j_2}$ such that $\ell\big(\wh{c}_{j_1},\wh{c}_{j_2};\wh{\sigma}_{j_1j_2}\big)=\wh{P}_{j_1j_2}=\sum_{i=1}^{n}I\big(Y_{ij_1}=Y_{ij_2}=1\big)/n$. Nevertheless, whether such an estimator $\wh{\sigma}_{j_1j_2}$ uniquely exists does not seem immediately clear. To address this concern, we need to gain more understandings of $\ell\big(\wh{c}_{j_1},\wh{c}_{j_2};\sigma_{j_1j_2}\big)$. 

Note that $\ell\big(\wh{c}_{j_1},\wh{c}_{j_2};\sigma_{j_1j_2}\big)$ is an upper bivariate normal tail probability. It can be verified that
\begin{equation}\label{equ:l_dot}
	\dot{\ell}\big(\wh{c}_{j_1},\wh{c}_{j_2};\sigma_{j_1j_2}\big)=\dfrac{\partial \ell\big(\wh{c}_{j_1},\wh{c}_{j_2};\sigma_{j_1j_2}\big)}{\partial\sigma_{j_1j_2}}=\dfrac{1}{\sqrt{1-\sigma_{j_1j_2}^2}}\phi\Bigg\{\dfrac{\wh{c}_{j_1}-\sigma_{j_1j_2}\wh{c}_{j_2}}{\sqrt{1-\sigma_{j_1j_2}^2}}\Bigg\}\phi\big(\wh{c}_{j_2}\big),
\end{equation}
where $\phi(\cdot)$ stands for the probability density function of a standard normal distribution \citep{sheppard1900,drezner1990}.  The verification details are given in \sc{Part 1}\normalfont$\ $of Appendix B.1. Obviously, we have $\dot{\ell}\big(\wh{c}_{j_1},\wh{c}_{j_2};\sigma_{j_1j_2}\big)> 0$ for $ \vert\sigma_{j_1j_2}\vert< 1$. As a consequence, $\ell\big(\wh{c}_{j_1},\wh{c}_{j_2};\sigma_{j_1j_2}\big)$ is a monotonically increasing function of $\sigma_{j_1j_2}$. By \cite{kotz2004}, we know that $\ell\big(\wh{c}_{j_1},\wh{c}_{j_2};1\big)=1-\Phi\big(\max\big(\wh{c}_{j_1},\wh{c}_{j_2}\big)\big)=1-\max\big\{\Phi\big(\wh{c}_{j_1}\big),\Phi\big(\wh{c}_{j_2}\big)\big\}$ and $\ell\big(\wh{c}_{j_1},\wh{c}_{j_2};-1\big)=1-\Phi\big(\wh{c}_{j_1}\big)-\Phi\big(\wh{c}_{j_2}\big)$ if $\wh{c}_{j_1}+\wh{c}_{j_2}\leq 0$, and if 0 otherwise. It can be also verified that $\ell\big(\wh{c}_{j_1},\wh{c}_{j_2};-1\big)<\ell\big(\wh{c}_{j_1},\wh{c}_{j_2};\sigma_{j_1j_2}\big)<\ell\big(\wh{c}_{j_1},\wh{c}_{j_2};1\big)$. The verification details are given in \sc{Part 2}\normalfont$\ $of Appendix B.1. Therefore, an inverse function of $P_{j_1j_2}^*=\ell\big(\wh{c}_{j_1},\wh{c}_{j_2};\sigma_{j_1j_2}\big)$ can be defined as $\ell^{-1}\big(\wh{c}_{j_1},\wh{c}_{j_2};P_{j_1j_2}^*\big)$. It happens that $\ell\big(\wh{c}_{j_1},\wh{c}_{j_2};-1\big)<\wh{P}_{j_1j_2}<\ell\big(\wh{c}_{j_1},\wh{c}_{j_2};1\big)$. Then, we can apply the inverse function $\ell^{-1}(\cdot) $ on $\wh{P}_{j_1j_2}$ so that a simple moment estimator can be obtained for $\sigma_{j_1j_2}$ as $\wh{\sigma}_{j_1j_2}=\ell^{-1}\big(\wh{c}_{j_1},\wh{c}_{j_2};\wh{P}_{j_1j_2}\big)$. 
Write $\theta_{j_1j_2}=\big(\theta_{j_1j_2}^1,\theta_{j_1j_2}^2,\theta_{j_1j_2}^3\big)^{\top}=\big(c_{j_1},c_{j_2},P_{j_1j_2}\big)^{\top}\in\mR^3$. Define $g\big(\theta_{j_1j_2}\big)=\ell^{-1}\big(c_{j_1},$ $c_{j_2};P_{j_1j_2}\big)$. Let $ \dot{g}\big(\theta_{j_1j_2}\big)=\partial g\big(\theta_{j_1j_2}\big)/$  $\partial\theta_{j_1j_2}\in\mR^3$ and $\ddot{g}\big(\theta_{j_1j_2}\big)=\partial^2 g\big(\theta_{j_1j_2}\big)/\partial\theta_{j_1j_2}\partial\theta_{j_1j_2}^{\top}\in\mR^{3\times 3}$ be the 1st and 2nd order partial derivatives of $ g\big(\theta_{j_1j_2}\big) $ with respect to $\theta_{j_1j_2}$, respectively. Let $A$ be an arbitrary matrix with dimension $M\times N$. Define its norm as $\Vert A\Vert=\lambda_{\max}^{1/2}(A^{\top}A)=\lambda_{\max}^{1/2}(AA^{\top})$, where $\lambda_{\max}(B)$ stands for the maximal eigenvalue of an arbitrary symmetric matrix $B$. To investigate the theoretical property of $\wh{\sigma}_{j_1j_2}$, the following technical conditions are necessarily needed. 
	\begin{enumerate}\setlength{\itemsep}{-2pt}
		\item[\textbf{(C1)}] \sc{(Univariate probability) }\normalfont Assume that there exist two positive constants $0<P_{\min}^{(1)}\leq P_{\max}^{(1)}<1$, such that $ P_{\min}^{(1)}\leq\min_{1\leq j\leq p}P_{j}\leq\max_{1\leq j\leq p}P_{j}\leq P_{\max}^{(1)}$. \label{con1} 
		 \item[\textbf{(C2)}] \sc{(Bivariate probability) }\normalfont Assume that there exist two positive constants $0<P_{\min}^{(2)}\leq P_{\max}^{(2)}<1$, such that $ P_{\min}^{(2)}\leq\min_{1\leq j_1,j_2\leq p}P_{j_1j_2}\leq\max_{1\leq j_1,j_2\leq p}P_{j_1j_2}\leq P_{\max}^{(2)}$. \label{con2}
	\end{enumerate}
\noindent Note that conditions (C1) and (C2) are standard regularity conditions for the true parameters, which are also used in the literature \citep{kotz2004,bartholomew2011latent}. Then, the asymptomatic behaviour of $\wh{\sigma}_{j_1j_2}$ is summarized in Theorem \ref{thm1}, which is proved in Appendix A.1. By Theorem \ref{thm1}, we know that $\wh{\sigma}_{j_1j_2}-\dot{g}\big(\theta_{j_1j_2}\big)^{\top}\big(\wh{\theta}_{j_1j_2}-\theta_{j_1j_2}\big) $ is extremely close to $\sigma_{j_1j_2}$ uniformly over $ 1\leq j_1\neq j_2\leq p$.
\begin{theorem}\label{thm1}
	Assume the technical conditions (C1) and (C2) hold. Then, for an arbitrary constant $ \epsilon>0$, there exist some sufficiently large but fixed constants $ M_{\epsilon}>0$ and $ M_{\epsilon}^g>0$ such that
	\begin{equation}\label{2.3}
		P\bigg\{\max\limits_{1\leq j_1\neq j_2\leq p}\Big\vert \wh{\sigma}_{j_1j_2}-\sigma_{j_1j_2}-\dot{g}\big(\theta_{j_1j_2}\big)^{\top}\big(\wh{\theta}_{j_1j_2}-\theta_{j_1j_2}\big)\Big\vert>\epsilon\bigg\}\leq 40p^2\exp\Big(-C_{\epsilon}^*n\Big),
	\end{equation}
	where $C_{\epsilon}^*=\min\Big\{2\epsilon^2/\big(9M_{\epsilon}^2+4M_{\epsilon}\epsilon\big),4\epsilon\big/\big(9M_{\epsilon}^2M_{\epsilon}^g+4M_{\epsilon}\sqrt{2M_{\epsilon}^g\epsilon}\big)\Big\}$.
\end{theorem}
\subsection{A Latent Factor Model}

We next consider how to impose a factor model structure on $\Sigma$, the covariance matrix of $e_i$. Following the classical idea of factor modelling \citep{1956statistical,fan2008}, we can impose a low-dimensional factor model structure on $\Sigma$. This can be done by assuming for $e_i$ a factor structure as
\begin{equation}\label{2.4}
	e_i = BZ_i+\varepsilon_i,
\end{equation}
where $Z_i=\big(Z_{i1},\ldots,Z_{id}\big)^{\top}\in\mR^d$ is a $d$-dimensional latent factor associated with the $i$-th subject, $ B=(\beta_{jk})\in\mR^{p\times d} $ is the loading matrix, and $\varepsilon_i=(\varepsilon_{i1},\ldots,\varepsilon_{ip})^{\top}\in\mR^p$ represents the information contained in $e_i$ but missed by $Z_i$. We further assume that $ E(\varepsilon_i)=0 $ and $\cov(\varepsilon_i) = D=\diag\big\{\tau_{1}^2,\ldots,\tau_{p}^2\big\}\in\mR^{p\times p}$ with  $\var(\varepsilon_{ij})=\tau_{j}^2$ for every $ 1\leq j\leq p$. We then have $ \Sigma=\cov(e_i)=B\cov(Z_i)B^\top+ D $. 

Obviously, the factor model (\ref{2.4}) is not uniquely identifiable. Write $\Sigma_Z=\cov(Z_i)\in\mR^{d\times d}$, which is assumed to be a positive definite matrix. We can then re-define $ Z_i\coloneqq \Sigma_Z^{-1/2}Z_i$ and $ B\coloneqq B\Sigma_Z^{1/2}$ so that model (\ref{2.4}) remains valid but $ \cov(Z_i)=I_d $, where $I_d $ is a $d\times d$ dimensional identity matrix. Unfortunately, even if we impose the constraint that $\Sigma_Z=I_d$, model (\ref{2.4}) still cannot be uniquely identified. To see this, write $\Sigma_B\coloneqq B^{\top}B/p\in\mR^{d\times d}$, which should be a positive definite matrix. Let $\lambda_k\big(\Sigma_B\big)$ be the $k$th largest eigenvalue of $\Sigma_B$ and $u_k\in\mR^d$ be the associated eigenvector for $1\leq k\leq d$. We can then construct an orthogonal matrix $U=(u_1,\ldots,u_d)\in\mR^{d\times d}$. Then, we can re-define $Z_i\coloneqq U^{\top}Z_i$ and $B\coloneqq BU$ such that model (\ref{2.4}) remains valid but $\Sigma_B=\diag\big\{\lambda_1\big(\Sigma_B\big),\ldots,\lambda_d\big(\Sigma_B\big)\big\}\in\mR^{d\times d}$ is a diagonal matrix. Therefore, the two assumptions $ \cov(Z_i)=I_d$ and $\Sigma_B=\diag\big\{\lambda_1\big(\Sigma_B\big),\ldots,\lambda_d\big(\Sigma_B\big)\big\}\in\mR^{d\times d} $ are conducted as follows.

Note that $ \Sigma=BB^\top+ D$. If we follow the existing literature \citep{1973estimation} and further assume that $ D=\tau_D^2I_p $ for some $\tau_D^2>0$ (i.e., $ \tau_{j}^2=\tau_D^2$ for every $ 1\leq j\leq p$), we then have $\beta_k/\sqrt{p\lambda_k\big(\Sigma_B\big)}$ with $ \beta_k=(\beta_{1k},\ldots,\beta_{pk})^{\top}\in\mR^p$ is a unit-length eigenvector of $\Sigma$ associated with the eigenvalue $p\lambda_k\big(\Sigma_B\big)+\tau_D^2$ for every $1\leq k\leq d$. Then, we can uniquely identify $\beta_k$ by assuming $\lambda_1\big(\Sigma_B\big)>\lambda_2\big(\Sigma_B\big)>\ldots>\lambda_d\big(\Sigma_B\big)$ and $ \beta_{jk(j)}>0$ for every $ 1\leq j\leq p$, where $k(j)=\argmax_k\big|\beta_{jk}\big|$. Since $ B=(\beta_1,\ldots,\beta_d)\in\mR^{p\times d}$, then $B$ can also be uniquely identified. Unfortunately, the assumption $ D=\tau_D^2I_p$ can hardly be satisfied since we are imposing a standard probit model (\ref{2.1}) on $ e_{ij} $. Therefore, we have $\var(e_{ij})=1 $ and $ \cov(e_{ij_1},e_{ij_2})=\sigma_{j_1j_2}$. This implies that $\tau_{j}^2+\Vert b_j \Vert^2=1$ with $ b_j=\big(\beta_{j1},\ldots,\beta_{jd}\big)^{\top}\in\mR^d$ and $ \sigma_{j_1j_2}=b_{j_1}^{\top}b_{j_2}$, respectively. Accordingly, it is impossible to have $ D=\tau_D^2I_p $ for some $ \tau_D^2>0$, unless we have $\Vert b_j\Vert^2=C_b$ for some constant $ C_b>0$. The consequence is that $ \beta_k/\sqrt{p\lambda_k\big(\Sigma_B\big)}s $ are extremely unlikely to be the eigenvectors of $\Sigma$. Therefore, if we stick with the traditional notion of identifiability, the eigenvector analysis of $ \Sigma $ cannot help much either. 

\subsection{Asymptotic Identifiability and Consistent Estimation}

To estimate the factor loading matrix $ B $ in model (\ref{2.4}), it is crucial to ensure its identifiability. Unfortunately, as mentioned before, the traditional notion of identifiability developed for problems with fixed dimensions is insufficient for high-dimensional problems with a diverging number of features. Therefore, a new type of identifiability concept needs to be developed. In this regard, we propose a novel notion of asymptotic identifiability in subspace. Specifically, instead of identifying $ B$ directly, we aim to identify the linear subspace spanned by the column vectors of $ B $, denoted as $ \mS(B)$. This idea has been extensively used in the literature of sufficient dimension reduction \citep{1991sliced,2009likelihood}. However, identifying $\mS(B)$ is also theoretically challenging. This is because we are dealing with a situation with a diverging feature dimension $p\to\infty$. In other words, $ \mS(B) $ as a parameter of interest is asymptotically varying as $ p\to\infty $. Therefore, we cannot study its identifiability in the same way as for the traditional situation with a fixed $ p $. Instead, we should study the identifiability issue dynamically with $ p\to\infty $. That leads to the idea of asymptotic identifiability. 

To establish a rigorous definition of the asymptotic identifiability, we define $ \Sigma_d=(v_1,\ldots,v_d)\in\mR^{p\times d} $, where $ v_k\in\mR^p\  (1\leq k\leq d)$ is the eigenvector associated with the $k$-th largest eigenvalue of $\Sigma$. Then, we have $ \Sigma_d^{\top}\Sigma_d=I_d$. To measure the discrepancy between $\mS(B) $ and $ \mS(\Sigma_d)$, we consider $\mD\big(B,\Sigma_d\big)=\tr\big(H_B-H_{\Sigma_d}\big)^2$ \citep{li2005,wang2008}. Specifically, $H_B=B(B^{\top}B)^{-1}B^{\top}$ and $H_{\Sigma_d}=\Sigma_d(\Sigma_d^{\top}\Sigma_d)^{-1}\Sigma_d^{\top}=\Sigma_d\Sigma_d^{\top}$ are the projection matrices onto the linear subspaces $\mS(B) $ and $ \mS(\Sigma_d)$, respectively. Note that $\mD\big(\cdot,\cdot\big)$ is symmetric in its arguments and thus $ \mD\big(B,\Sigma_d\big)=\mD\big(\Sigma_d,B\big)$. In an ideal scenario with $\mD(B,\Sigma_d)=0$, we should have $ \mS(B)=\mS(\Sigma_d)$. However, under our high-dimensional model setup, we can hardly have $ \mD(B,\Sigma_d)=0 $, unless $ D=\tau_D^2I_p$ for some $ \tau_D^2>0$. Instead, we might have that $ \mD(B,\Sigma_d)\approx0 $ in a carefully defined asymptotic sense. Specifically, we say that $ \mS(B)$ can be asymptotically identified if one can show that $\mD\big(B,\Sigma_d\big)\to 0 $ as $ p\to\infty $. In this case, the two linear subspaces $ \mS(B)$ and $\mS(\Sigma_d)$ should be asymptotically close to each other as $ p\to\infty $. 

Once $\mS(B)$ can be asymptotically identified by $\mS(\Sigma_d)$, we consider how to estimate it consistently. By the asymptotic identifiability, a natural estimator should be $\mS(\wh{\Sigma}_d)$, where $\wh{\Sigma}_d=(\wh{v}_1,\ldots,\wh{v}_d)\in\mR^{p\times d} $ and $ \wh{v}_k\in\mR^p\  (1\leq k\leq d)$ is the eigenvector associated with the $k$-th largest eigenvalue of $\wh{\Sigma}$. Recall that $ \wh{\Sigma} $ is the consistent estimator of $ \Sigma $ developed in Section 2.1. Subsequently, the discrepancy between $\mS(B)$ and $\mS(\wh{\Sigma}_d)$ can be evaluated by $\mD\big(B,\wh{\Sigma}_d\big)$. We then make the following conditions. 

\begin{enumerate}
	\item[\textbf{(C3)}] \label{con3} \sc{(Variance of errors) }\normalfont Assume that there exist two positive constants $0< \tau_{\min}\leq\tau_{\max}<1 $, such that $ \tau_{\min}\leq\min_{1\leq j\leq p}\tau_j\leq \max_{1\leq j\leq p}\tau_{j}\leq \tau_{\max}$.
	\item[\textbf{(C4)}] \label{con4} \sc{(Divergence speed of $p$) }\normalfont Assume that there exist some positive constants $\xi$ and $\rho$, such that $ p\leq \rho n^{\xi}$ and $\xi<1/2$.
\end{enumerate}

\noindent To asymptotically identify and consistently estimate $\mS(B)$, conditions (C3) and (C4) are sufficient. Specifically, condition (C3) is a standard regularity condition for the true $\tau_js$ \citep{kotz2004,bartholomew2011latent}. By condition (C4), we allow the feature dimension $p$ to diverge with sample size but require it to be much less than the sample size $n$. We then develop the following theorem, whose proof is given in Appendix A.2. By Theorem \ref{thm2}, we know that $\mS(B)$ can be asymptotically identified by $\mS(\Sigma_d)$ in the sense of $\mD\big(B,\Sigma_d\big)\to 0 $ as $ p\to\infty $. Moreover, $\mS(B)$ can be consistently estimated by $\mS(\wh{\Sigma}_d)$ in the sense of $\mD\big(B,\wh{\Sigma}_d\big)\to 0 $ as $ n\to\infty $.
\begin{theorem}\label{thm2}
	Assume the technical conditions (C1)-(C4) hold. Then, we should have (1) $ \mD\big(B,\Sigma_d\big)\to 0 $ as $ p\to\infty $ and (2) $ \mD\big(B,\wh{\Sigma}_d\big)\xrightarrow{p}0$ as $ n\to\infty $.
\end{theorem}
\subsection{Latent Factor Estimation}

To estimate the latent factor $Z_i$ in (\ref{2.4}), we must first estimate $\tau_j^2=\var(\varepsilon_{ij})$ for every $1\leq j\leq p$. Recall that $\varepsilon_i$ represents the information contained in $e_i$ but missed by $Z_i$. Let $Q_B=I_p-H_B$ be the projection matrix onto the orthogonal subspace of $\mS(B)$, denoted as $\mS^{\perp}(B)$. Note that $\Sigma=BB^{\top}+D$. Therefore, we can approximate $D=\diag\big\{\tau_1^2,\ldots,\tau_p^2\big\}$ by projecting $\Sigma$ onto $\mS^{\perp}(B)$, resulting in $Q_B\Sigma Q_B^{\top}=Q_BDQ_B^{\top}$ so that $\diag\big\{Q_B\Sigma Q_B^{\top}\big\}=\diag\big\{Q_BDQ_B^{\top}\big\}\approx D$. By Theorems \ref{thm1} and \ref{thm2}, we know that $\wh{\Sigma}$ and $\mS(\wh{\Sigma}_d)$ are consistent estimators for $\Sigma$ and $\mS(B) $, respectively. This implies that $Q_{\wh{\Sigma}_d}=I_p-H_{\wh{\Sigma}_d}$ is also a consistent estimator for $Q_B$ in some sense. Then, a natural estimator for $D$ can be constructed as $\wh{D}=\diag\big\{Q_{\wh{\Sigma}_d}\wh{\Sigma}Q_{\wh{\Sigma}_d}^{\top}\big\}=\diag\big\{\wh{\tau}_1^2,\ldots,\wh{\tau}_p^2\big\}$. Its theoretical property is then summarized in Theorem \ref{thm3}, whose proof is given in Appendix A.3. By Theorem \ref{thm3}, we know that $\wh{\tau}_j^2$ converges to $\tau_j^2$ consistently on average over $j$.
\begin{theorem}\label{thm3}
	Assume the technical conditions (C1)-(C4) hold, we then should have $ p^{-1}\sum_{j=1}^{p}\big\vert\wh{\tau}_j^2-\tau_j^2\big\vert\xrightarrow{p}0$ as $n\to\infty$. 
\end{theorem}

After obtaining consistent estimators for $\tau_j^2s$, we can proceed to estimate the latent factor $Z_i$ in (\ref{2.4}). By Theorem \ref{thm2}, we know that $\mS(\wh{\Sigma}_d)$ is a consistent estimator for $\mS(B)$. Therefore, we can use $\wh{\Sigma}_d$ as an estimator for one particular set of the basis for $\mS(B)$ and construct $\wh{B}=\wh{\Sigma}_d\wh{\Lambda}_d^{1/2}$ with $\wh{\Lambda}_d^{1/2}=\diag\big\{\lambda_1^{1/2}\big(\wh{\Sigma}\big),\ldots,\lambda_d^{1/2}\big(\wh{\Sigma}\big)\big\}$. Next, we treat $\wh{B}$ as if it were the true factor loading matrix $B$ and specify a set of probit regression models for every sample $i$ as
\begin{equation}\label{2.5}
	P\Big(Y_{ij}=1\big\vert Z_i\Big)=\Phi\Big\{\Big(b_j^{\top}Z_i-c_j\Big)/\tau_j\Big\}\approx \Phi\Big\{\Big(\wh{b}_j^{\top}Z_i-\wh{c}_j\Big)/\wh{\tau}_j\Big\},
\end{equation}
where $ b_j=\big(\beta_{j1},\ldots,\beta_{jd}\big)^{\top}\in\mR^d$, $ \wh{B}=(\wh{\beta}_{jk})\in\mR^{p\times d} $, $ \wh{b}_j=\big(\wh{\beta}_{j1},\ldots,\wh{\beta}_{jd}\big)^{\top}\in\mR^d$, $\wh{c}_j=-\Phi^{-1}\big(\wh{P}_j\big)$ and $ 1\leq j\leq p $. Then, by treating $Z_i$ as if it were an unknown regression coefficient in (\ref{2.5}), a log-likelihood function for $Z_i$ can be constructed as
\begin{equation}\label{2.6}
		\mL_{\wh{B}}^{(i)}\big(Z_i\big)=p^{-1}\sum_{j=1}^p\Bigg(Y_{ij}\log\Phi\Big\{\Big(\wh{b}_j^{\top}Z_i-\wh{c}_j\Big)/\wh{\tau}_j\Big\}+\Big(1-Y_{ij}\Big)\log\bigg[1-\Phi\Big\{\Big(\wh{b}_j^{\top}Z_i-\wh{c}_j\Big)/\wh{\tau}_j\Big\}\bigg]\Bigg).
\end{equation}
Practically, the log-likelihood function $\mL_{\wh{B}}^{(i)}\big(Z_i\big) $ could be very unstable and even sensitive to $Z_i$ if there exists some $\wh{\tau}_j^2$ extremely close to 0. To fix this problem, we further restrict the log-likelihood function to be a more stable one as
\begin{equation}\label{2.7}
		\mL_{\wh{B},\tau}^{(i)}\big(Z_i\big)=p^{-1}\sum_{j=1}^pI\Big(\wh{\tau}_j>\tau\Big)\Bigg(Y_{ij}\log\Phi\Big\{\Big(\wh{b}_j^{\top}Z_i-\wh{c}_j\Big)/\wh{\tau}_j\Big\}+\Big(1-Y_{ij}\Big)\log\bigg[1-\Phi\Big\{\Big(\wh{b}_j^{\top}Z_i-\wh{c}_j\Big)/\wh{\tau}_j\Big\}\bigg]\Bigg),
\end{equation}
where $0<\tau<\tau_{\min}$ is some pre-specified and fixed constant. For example, it can be practically decided as a number so that $m$\% (e.g., $m=90$) of $j$-components of the log-likelihood function vector can be included in (\ref{2.7}). Then, an estimator for $ Z_i $ can be obtained as $ \wh{Z}_i=\argmax_{Z_i}\mL_{\wh{B},\tau}^{(i)}\big(Z_i\big) $. To establish the property of $\wh{Z}_i$, the following condition is required.
\begin{spacing}{1.2}
	\begin{enumerate}
		\item[\textbf{(C5)}] \label{con5} \sc{(Latent factors) }\normalfont We assume that $\max_{1\leq i\leq n}\Vert Z_i\Vert\leq Z_{\max} $ for some sufficiently large constant $ Z_{\max}>0$, where $ \Vert\cdot\Vert $ denotes the standard $L_2$ norm.
	\end{enumerate}
\end{spacing}
\noindent Condition (C5) assumes that the latent factor is bounded. A similar condition can be found in \cite{1978contributions} and \cite{2009statistical}. Based on the above conditions, we have the following theorem about the asymptotic property of the associated estimator. The proof details are presented in Appendix A.4. By Theorem \ref{thm4}, we know that $\wh{B}\wh{Z}_i$ is a consistent estimator for $ BZ_i$ as a whole. 
\begin{theorem}\label{thm4}
	Assume the technical conditions (C1)-(C5) hold, we should then have $ p^{-1/2}\big\Vert\wh{B}\wh{Z}_i-BZ_i\big\Vert \xrightarrow{p}0$ as $n\to\infty$. 
\end{theorem}
\section{Numerical Studies\label{sec:3}}

\subsection{The Simulation Setup}

We present several simulation studies to demonstrate the finite sample performance of the latent factor model for the high-dimensional binary data. Specifically, we aim to investigate the performance of the estimators proposed in the previous section. Those estimators are, respectively: (1) the covariance matrix estimator $\wh{\Sigma}=\big(\wh{\sigma}_{j_1j_2}\big)\in\mR^{p\times p}$, (2) the loading matrix subspace estimator $\mS\big(\wh{\Sigma}_d\big)$, and (3) the latent factor estimator $\wh{Z}_i\in\mR^d$. Following \cite{wang2012}, we consider a total of three different factor structures with $d=1,2$ or $3$. For the entire simulation experiment, we set the dimension of the binary vector to be $p=50$, $80$ or $100$. The sample sizes are set to be  $n=4,000+2,000r$ with $r=0,1,\ldots,5$ as long as $n\geq p^2$. 

Once those dimensions (i.e., $d$, $p$ and $n$) are given, we next consider how to generate other parameters appropriately. Following \cite{fan2008sure}, the latent factor $Z_i\in\mR^d$ is generated from a $d$-dimensional normal distribution with a mean of 0 and covariance $I_d$. Different $\beta_{jk}s$ are generated from a uniform distribution between -1 and 1. Then, we obtain $B=\big(\beta_{jk}\big)\in\mR^{p\times d}$.  For the variance of the noise term, $\tau_j^2s$ are generated from a uniform distribution between 0.2 and 0.8. Thereafter, $\varepsilon_{ij}$ is generated from the normal distribution with a mean of 0 and variance $ \tau_j^2$. This leads to the continuous term $e_i$ according to (\ref{2.4}). Next, $c_js$ in the probit model (\ref{2.1}) are generated from a uniform distribution between -1 and 1. Thus, the high-dimensional binary data $Y=\big(Y_{ij}\big)\in\mR^{n\times p}$ is obtained according to the model (\ref{2.1}).
\subsection{The Covariance Matrix Estimator}

\begin{figure}[b]
	\centering
	\includegraphics[width=6in]{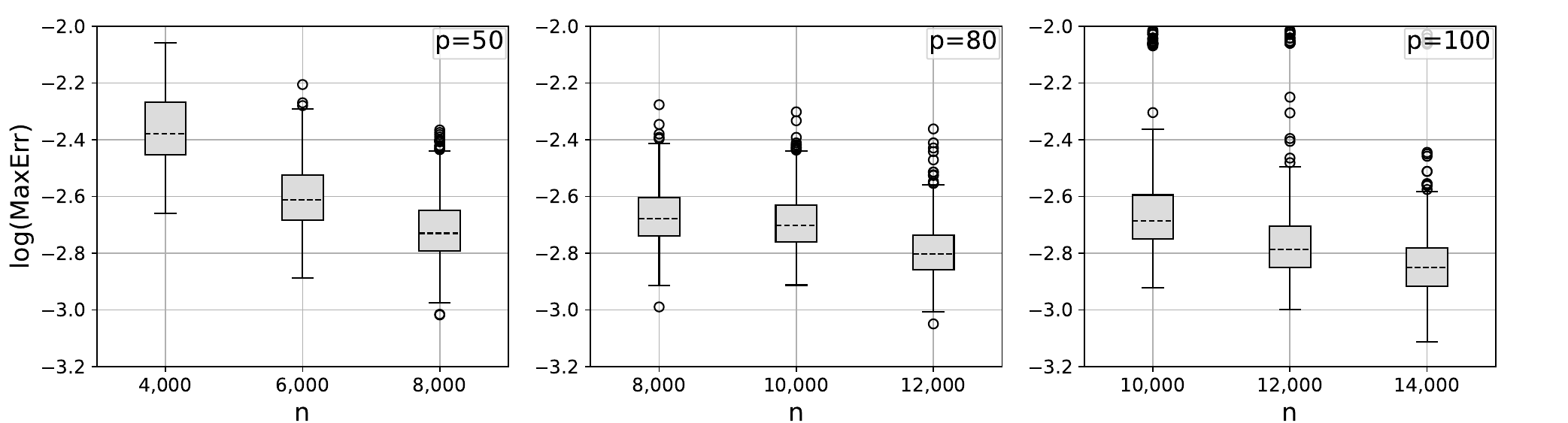}
	\caption{The log(MaxErr) values for the covariance matrix estimator $\wh{\Sigma}=\big(\wh{\sigma}_{j_1j_2}\big)\in\mR^{p\times p}$ with $d=2$. Different panels correspond to different feature dimensions: $p=50$ (the left), $80$ (the middle), and $100$ (the right). For a given panel, different boxplots correspond to different sample sizes.}
	\label{fig:sigma}
\end{figure}

We start with the covariance matrix estimator $\wh{\Sigma}=\big(\wh{\sigma}_{j_1j_2}\big)\in\mR^{p\times p}$. Recall that $\wh{\sigma}_{j_1j_2}=\ell^{-1}\big(\wh{c}_{j_1},\wh{c}_{j_1};\wh{P}_{j_1j_2}\big)$ and $\wh{c}_j=-\Phi^{-1}\big(\wh{P}_j\big)$. To evaluate the finite sample performance of $\wh{\sigma}_{j_1j_2}$, we should first study $\wh{P}_j$ and $\wh{P}_{j_1j_2}$ separately. Since they are both simple moment estimators, we only present the simulation results of $ \wh{\sigma}_{j_1j_2}s$ for illustration purpose. For a given sample size $n$ and feature dimension $p$, the experiment is randomly replicated for a total of $R = 1,000$ times. We use  $ \wh{\sigma}_{j_1j_2}^{(r)}$ to represent one particular estimator obtained in the $r$-th replication ($1\leq r \leq R$). Then, the estimation error (Err) can be evaluated as Err$_{j_1j_2}^{(r)}=\big\vert\wh{\sigma}_{j_1j_2}^{(r)}-\sigma_{j_1j_2}\big\vert$ for every $\wh{\sigma}_{j_1j_2}^{(r)}$ and its maximum error (MaxErr) over $j_1$ and $j_2$ is given by $\text{MaxErr}^{(r)}=\max_{j_1,j_2}\text{Err}_{j_1j_2}^{(r)}$.  We fix the factor dimension $d=2$. Those MaxErr values for different $(n,p)$ combinations are then log-transformed and box-plotted in Figure \ref{fig:sigma}. By Figure \ref{fig:sigma}, we can see that given $p$, large sample sizes always lead to smaller MaxErr values. This confirms that $\wh{\sigma}_{j_1j_2} $ is uniformly consistent for $\sigma_{j_1j_2}$ over $1\leq j_1\neq j_2\leq p$, which is in line with the result in Theorem \ref{thm1}. Other factor dimensions are also studied. The simulation results are qualitatively similar to those of $d=2$. Therefore, they are omitted to save space.
\subsection{The Loading Matrix Subspace Estimator}

\begin{figure}[]
	\centering
	\includegraphics[width=6in]{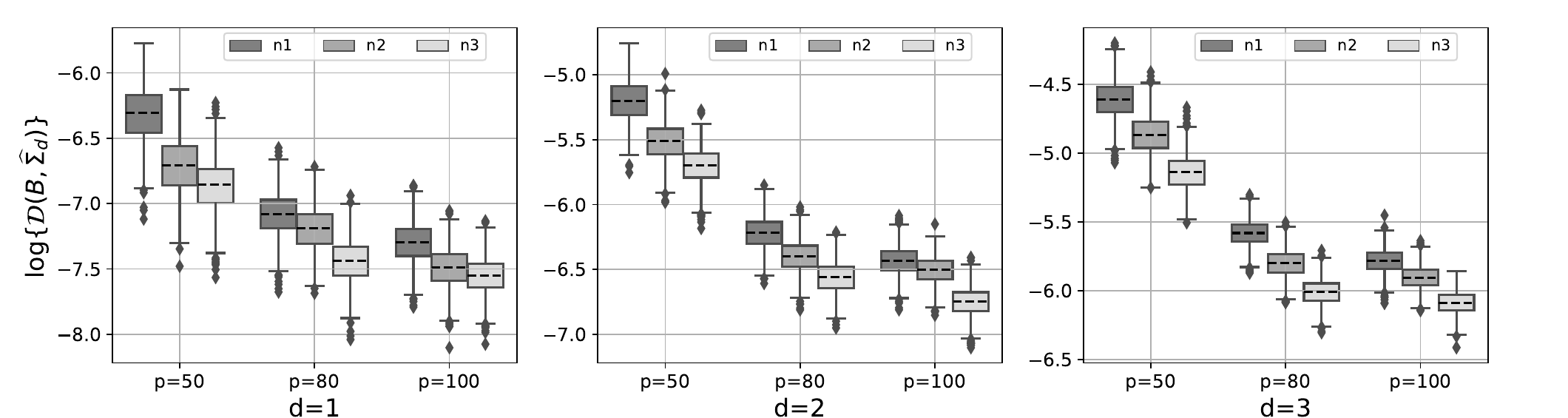}
	\caption{The log$\big\{\mD\big(B,\wh{\Sigma}_d\big)\big\}$ values for the factor loading matrix $B$ and its estimator $\wh{\Sigma}_d$. Different panels correspond to different latent factor dimensions: $d=1$ (the left), 2 (the middle), and 3 (the right). For a given panel, different groups correspond to different feature dimensions with $p=50$, $80$ and $100$, respectively. For a given group, the lighter the colour of the box, the larger the sample size (e.g., $n_1=4,000$, $n_2=6,000$ and $n_3=8,000$ for $p=50$).}
	\label{fig:HB}
\end{figure}

We next study the finite sample performance of  $\mS\big(\wh{\Sigma}_d\big)$, as an estimator for $\mS\big(B\big)$. Recall that the measure for the discrepancy between $ \mS\big(B\big)$ and $\mS\big(\wh{\Sigma}_d\big) $ is evaluated by $\mD\big(B,\wh{\Sigma}_d\big)=\tr\big(H_B-H_{\wh{\Sigma}_d}\big)^2$. Recall that we have $R = 1,000$ random replications. This leads to a total of $R$ randomly replicated $\mD\big(B,\wh{\Sigma}_d\big)$ values for $d=1,2$ and $3$. We present the simulation results of all factor dimensions. Those $\mD\big(B,\wh{\Sigma}_d\big)$  values are then log-transformed and box-plotted in Figure \ref{fig:HB}. By Figure \ref{fig:HB}, we obtain the following two interesting findings. First, we find that as the sample size $n$ increases, the discrepancy between $\mS\big(B\big)$ and $ \mS\big(\wh{\Sigma}_d\big)$ decreases for each combination of $(p,d)$. This implies that $\mS\big(\wh{\Sigma}_d\big)$ should be a consistent estimator for $\mS\big(B\big)$. Moreover, we can also deduce that the larger the feature dimension $p$, the smaller the discrepancy measure $ \mD\big(B,\wh{\Sigma}_d\big)$ values for every case. This provides some numerical evidence of the 1st conclusion of Theorem \ref{thm2}, that is the asymptotic identifiability of $\mS(B)$.
\subsection{The Latent Factor Extraction}

\begin{figure}[]
	\centering
	\includegraphics[width=3in]{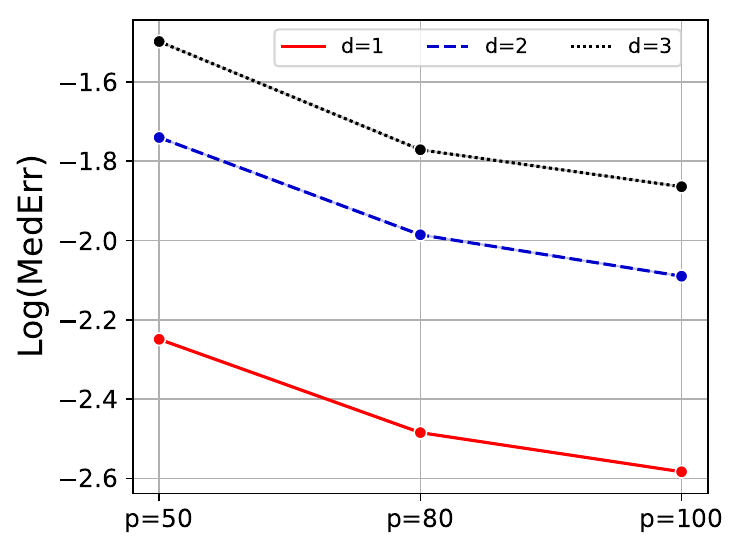}
	\caption{The log(MedErr) values for $\wh{B}\wh{Z}_i$. Different colored lines correspond to different latent factor dimensions: $d=1$ (the red line), 2 (the blue line), and 3 (the black line). For a given line, different points correspond to different feature dimensions with $p=50$, $80$ and $100$, respectively. }
	\label{fig:BZ}
\end{figure}

Lastly, we are extremely interested in the finite sample performance of the latent factor estimator $\wh{Z}_i$. Unfortunately, $\wh{Z}_i$ is not a consistent estimator for $Z_i$ since $Z_i$ is not identifiable. However, $BZ_i$ can be uniquely identified. Therefore, we focus on evaluating the estimation accuracy of $\wh{B}\wh{Z}_i$ as a whole. Similar to Section 3.2, we use $\wh{B}^{(r)}$ and $\wh{Z}_i^{(r)}$ to represent the estimators obtained in the $r$-th replication. Then, the estimation error can be evaluated by $ \text{Err}_i^{(r)} = p^{-1/2}\big\Vert\wh{B}^{(r)}\wh{Z}_i^{(r)}-BZ_i\big\Vert$ for every $1\leq i\leq n$ and $1\leq r\leq R$. We next use $\text{MedErr}^{(r)}$ to represent the median of those $\text{Err}_i^{(r)}$ values over $i$. Furthermore, the median of those $\text{MedErr}^{(r)}$ values is denoted as MedErr. Those MedErr values for different $(p,d)$ combinations are then log-transformed and plotted in Figure \ref{fig:BZ}. We find that larger feature dimensions always lead to smaller MedErr values. This is because larger feature dimensions provide more information about the latent factor and then more accurate estimator for the latent factor.

\subsection{A Real Data Example}

In this subsection, we present a case study to demonstrate the practical applications of the proposed methodology. Specifically, we consider the text descriptions of different categories of products obtained from JD.com (\url{https://www.jd.com}), which is one of the largest online retailers and marketplaces in China. The full dataset contains a total of 450,041 Chinese product descriptions. They belongs to three categories: computers (297,891), fresh food (132,174), and books (19,976). To obtain a representative sample, we randomly select 18,000 descriptions from each category, resulting in a final sample size of $n=54,000$. For each of the selected samples, we construct a high-dimensional binary feature vector $Y_i\in\mR^p$ with $p=200$. Specifically, we use a pre-specified set of keywords that are most relevant to the product descriptions. Then, we create a binary variable to indicate the presence of one particular keyword. This leads to a final high-dimensional binary data $Y = (Y_{ij})\in\mR^{n\times p}$ with $n=54,000$ and $p=200$.

\begin{figure}[]
	\centering
	\includegraphics[width=6in]{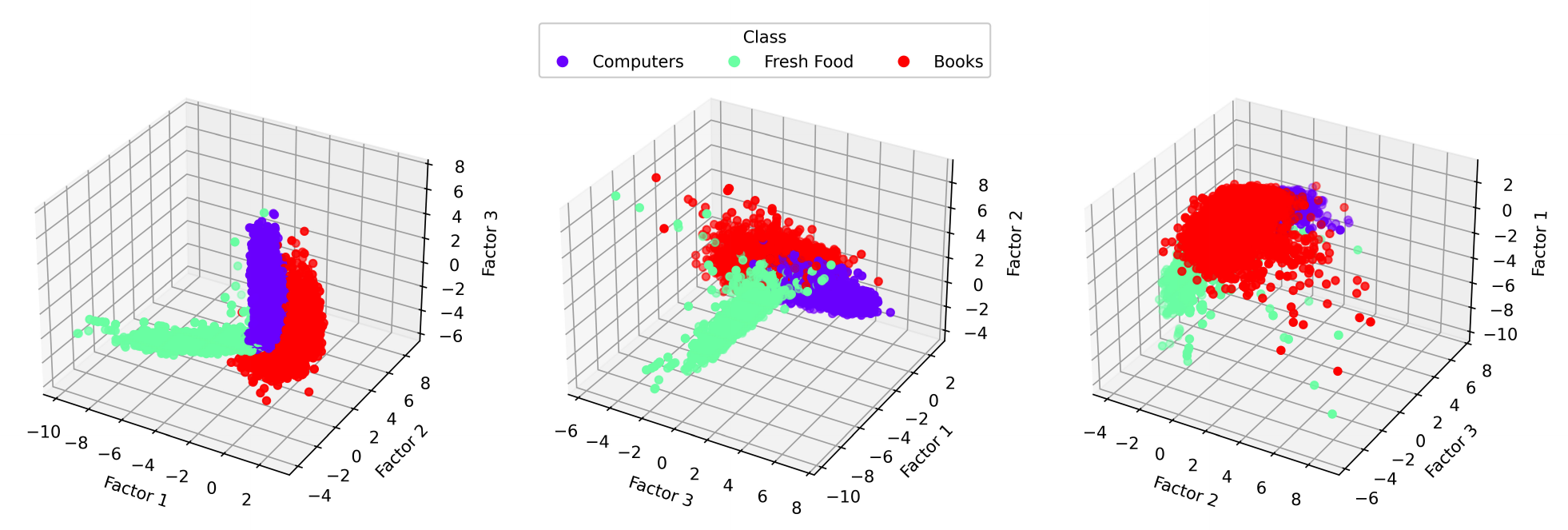}
	\caption{The extracted factors $\wh{Z}_is$ for $d=3$. Different panels correspond to different factor rankings: Factor 1, 2, 3 (the left panel), Factor 3, 1, 2 (the middle panel), and Factor 2, 3, 1 (the right panel). For a given panel, different colours represent different classes: computers (purple), fresh food (green), and books (red), respectively. }
	\label{fig:factor}
\end{figure}

To apply the proposed method to the high-dimensional binary data $Y=(Y_{ij})\in\mR^{n\times p}$, we should first determine the dimension of the latent factor. We consider several factor dimensions with $ d=1,2,3,4$ and $5$. Thereafter, the latent factor $\wh{Z}_i\in\mR^d$ can be computed for each sample $i$ and dimension $d$, as described in Section 2.4. Once the latent factors are obtained, we can evaluate the relationship between the true categories of the product descriptions and the extracted factors. Figure \ref{fig:factor} provides the details of the extracted factors for $d=3$ from three directions. It is evident that the extracted  factors are clustered into three distinct classes. This finding confirms the existence of the latent factor structure of the high-dimensional binary data $Y$ to some extent.

\section{Conclusions\label{sec:4}}

In this study, we develop a latent factor model for high-dimensional binary data analysis. A moment-based estimation method has been developed and the asymptotic properties of the resulting estimators have been carefully studied. To conclude this work, we discuss some interesting topics for future research. First, even though our current method allows the feature dimension to diverge with the sample size, we still require that the sample size be much larger than the feature dimension. Hence, it is worth exploring how to accommodate the ultrahigh dimensional binary features for factor analysis. Second, the distribution of the continuous latent variables in our model is assumed to admit a Gaussian distribution. This is an assumption that is obviously too restrictive. Then, how to relax this parametric assumption is another interesting direction for future exploration. Lastly, our current method is developed as a tool for exploratory analysis without a clearly defined response variable. However, high-dimensional binary features are commonly encountered in regression analysis with a specific response variable. Therefore, how to study such type of problems is also worth pursuing.

\section*{Acknowledgments}

The research of Rui Pan is supported by National Natural Science Foundation of China (NSFC, 11971504), the Disciplinary Funds and the Emerging Interdisciplinary Project of Central University of Finance and Economics. Hansheng Wang's research is partially supported by National Natural Science Foundation of China (12271012).

\section*{Appendix A. Proof of the Main Theoretical Results}
\subsection*{\textbf{Appendix A.1. Proof of Theorem 1}}

The theorem conclusion can be proved in a total of three steps. In the first step, we show that $ P\big\{\max_{j_1,j_2}\big\Vert \wh{\theta}_{j_1j_2}-\theta_{j_1j_2}\big\Vert>\epsilon\big\}\leq 4p^2\exp\big(-C_{\epsilon}^an\big)$ with $ C_{\epsilon}^a=2\epsilon^2/\big(9M_{\epsilon}^2+4M_{\epsilon}\epsilon\big)$ and $M_{\epsilon}=\max\Big\{\sup_{\tau\in\big(P_{\min}^{(1)}-\epsilon/3, P_{\max}^{(1)}+\epsilon/3\big)}\big\vert\phi\big(\Phi^{-1}(\tau)\big)\big\vert^{-1},1\Big\}$ for any given $\epsilon>0$. In the second step, we show that there exists a sufficiently large but fixed positive constant $M_{\epsilon}^g$ such that $ P\big\{\max_{j_1,j_2}\sup_{0\leq\eta\leq1}\big\Vert\ddot{g}\big(\wt{\theta}_{j_1j_2}\big)\big\Vert>M_{\epsilon}^g\big\}\leq 36p^2\exp\big(-C_{\epsilon}^an\big)$, where $\wt{\theta}_{j_1j_2}=\eta\wh{\theta}_{j_1j_2}+(1-\eta)\theta_{j_1j_2}$ for any $\eta\in[0,1]$. In the last step, we show that $P\big\{\max_{1\leq j_1\neq j_2\leq p}\big\vert \wh{\sigma}_{j_1j_2}-\sigma_{j_1j_2}-\dot{g}\big(\theta_{j_1j_2}\big)^{\top}\big(\wh{\theta}_{j_1j_2}-\theta_{j_1j_2}\big)\big\vert>\epsilon\big\}\leq 40p^2\exp\big(-C_{\epsilon}^*n\big)$, where $C_{\epsilon}^*=\min\big(C_{\epsilon}^a,C_{\epsilon}^b\big) $ and $ C_{\epsilon}^b=4\epsilon\big/\big(9M_{\epsilon}^2M_{\epsilon}^g+4M_{\epsilon}\sqrt{2M_{\epsilon}^g\epsilon}\big) $. The details are given below.

\sc{Step 1.}\normalfont$\ $We start with $\mQ_1= P\big\{\max_{j_1,j_2}\big\Vert \wh{\theta}_{j_1j_2}-\theta_{j_1j_2}\big\Vert>\epsilon\big\}$ for any given $\epsilon>0$. Recall that $\theta_{j_1j_2}=\big(-\Phi^{-1}(P_{j_1}),-\Phi^{-1}(P_{j_2}),P_{j_1j_2}\big)^{\top}\in\mR^3$. We then have $\max_{j_1,j_2}\big\Vert \wh{\theta}_{j_1j_2}-\theta_{j_1j_2}\big\Vert\leq\max_{j_1,j_2}\big\vert\wh{P}_{j_1j_2}-P_{j_1j_2}\big\vert+\max_{j_1,j_2}\big\vert\Phi^{-1}\big(\wh{P}_{j_1}\big)-\Phi^{-1}\big(P_{j_1}\big)\big\vert+\max_{j_1,j_2}\big\vert\Phi^{-1}\big(\wh{P}_{j_2}\big)-\Phi^{-1}\big(P_{j_2}\big)\big\vert$. Then, we can obtain that $\mQ_1\leq\mQ_1^a+\mQ_1^b+\mQ_1^c$, where $ \mQ_1^a=P\big\{\max_{j_1,j_2}\big\vert\wh{P}_{j_1j_2}$  $-P_{j_1j_2}\big\vert>\epsilon/3\big\}$, $ \mQ_1^b=P\big\{\max_{j_1}\big\vert\Phi^{-1}\big(\wh{P}_{j_1}\big)-\Phi^{-1}\big(P_{j_1}\big)\big\vert>\epsilon/3\big\}$ and $\mQ_1^c=P\big\{\max_{j_2}$  $\big\vert\Phi^{-1}\big(\wh{P}_{j_2}\big)-\Phi^{-1}\big(P_{j_2}\big)\big\vert>\epsilon/3\big\}$. To upper bound $ \mQ_1 $, it suffices to upper bound $ \mQ_1^a $, $ \mQ_1^b $ and $ \mQ_1^c $, respectively. Since the analysis of $ \mQ_1^b $ and $\mQ_1^c $ is nearly identical, we shall present the details for $ \mQ_1^a $ and $\mQ_1^b$. 

\sc{Step 1.1.}\normalfont$\ $We start with $\mQ_1^a$. Direct computation leads to $\mQ_1^a\leq \sum_{j_1,j_2}W_{j_1j_2}$, where  $W_{j_1j_2}=P\big\{\big\vert\wh{P}_{j_1j_2}-P_{j_1j_2}\big\vert>\epsilon/3\big\}$. Recall that $\wh{P}_{j_1j_2}=\sum_{i=1}^{n}I\big(Y_{ij_1}=Y_{ij_2}=1\big)/n$. Obviously, we have $0\leq I\big(Y_{ij_1}=Y_{ij_2}=1\big)\leq 1$ and then $0\leq \wh{P}_{j_1j_2}\leq 1$. We then compute  $E\big\{I\big(Y_{ij_1}=Y_{ij_2}=1\big)\big\}=P_{j_1j_2}$ and $\var\big\{I\big(Y_{ij_1}=Y_{ij_2}=1\big)\big\}=P_{j_1j_2}\big(1-P_{j_1j_2}\big)\leq 1/4$. Thus, the Bernstein's Inequality \citep{bernstein1926} can be applied as
\begin{equation}\label{b1}
	P\Big\{\big\vert\wh{P}_{j_1j_2}-P_{j_1j_2}\big\vert>\epsilon/3\Big\}\leq 2\exp\bigg\{-\dfrac{n\epsilon^2/18}{P_{j_1j_2}\big(1-P_{j_1j_2}\big)+\epsilon/9}\bigg\}.
\end{equation}
Subsequently, we have $W_{j_1j_2}\leq 2\exp\big\{-C_{\epsilon}n\big\}$, where $ C_{\epsilon}=2\epsilon^2/\big(9+4\epsilon\big)$. Thus, we can obtain that $ \mQ_1^a\leq 2p^2\exp\big(-C_{\epsilon}n\big)$.

\sc{Step 1.2.}\normalfont$\ $To study $\mQ_1^b=P\big\{\max_{j_1}\big\vert\Phi^{-1}\big(\wh{P}_{j_1}\big)-\Phi^{-1}\big(P_{j_1}\big)\big\vert>\epsilon/3\big\}$, we can conduct the Taylor's expansion about $\Phi^{-1}\big(\wh{P}_{j_1}\big)$ for $\wh{P}_{j_1}$ at $P_{j_1}$ as $\Phi^{-1}\big(\wh{P}_{j_1}\big)-\Phi^{-1}\big(P_{j_1}\big)=\dot{\Phi}^{-1}\big(\wt{P}_{j_1}\big)\big(\wh{P}_{j_1}-P_{j_1}\big)$, where $\dot{\Phi}^{-1}\big(P_{j_1}\big)=\partial\Phi^{-1}\big(P_{j_1}\big)/\partial P_{j_1}$ is the first-order partial derivative of $\Phi^{-1}\big(P_{j_1}\big)$ with respect to $ P_{j_1}$ and $\wt{P}_{j_1}=\eta_{j_1}\wh{P}_{j_1}+(1-\eta_{j_1})P_{j_1}$ for some $ \eta_{j_1}\in(0,1)$. Then, we have $ \max_{j_1}\big\vert\Phi^{-1}\big(\wh{P}_{j_1}\big)-\Phi^{-1}\big(P_{j_1}\big)\big\vert\leq \max_{j_1}\sup_{0<\eta<1}\big\vert\dot{\Phi}^{-1}\big(\wt{P}_{j_1}\big)\big\vert\max_{j_1}\big\vert\wh{P}_{j_1}$ $-P_{j_1}\big\vert$. Further, for an arbitrary large but fixed constant $M>0$, we have $\mQ_1^b\leq\mQ_1^{b_1}+\mQ_1^{b_2}$, where $\mQ_1^{b_1}=P\big\{\max_{j_1}\big\vert\wh{P}_{j_1}-P_{j_1}\big\vert>\epsilon/(3M)\big\}$ and $ \mQ_1^{b_2}=P\big\{\max_{j_1}\sup_{0<\eta<1}$ $\big\vert\dot{\Phi}^{-1}\big(\wt{P}_{j_1}\big)\big\vert>M\big\}$. Similar to \sc{Step 1.1,}\normalfont$\ $it can be verified that $\mQ_1^{b_1}\leq 2p\exp\big(-C_{\epsilon}^an\big)$ with $C_{\epsilon}^a=2\epsilon^2/\big(9M^2+4M\epsilon\big)$. Next, we should focus on $\mQ_1^{b_2}$.

Note that $\wt{P}_{j_1}=\eta_{j_1}\wh{P}_{j_1}+(1-\eta_{j_1})P_{j_1}$ for some $ \eta_{j_1}\in(0,1)$. We then have $\max_{j_1}\big\vert\wt{P}_{j_1}-P_{j_1}\big\vert\leq \max_{j_1}\big\vert\wh{P}_{j_1}-P_{j_1}\big\vert$. Therefore, we know that $ P\big\{\max_{j_1}\big\vert\wt{P}_{j_1}-P_{j_1}\big\vert>\epsilon/3\big\}\leq P\big\{\max_{j_1}\big\vert\wh{P}_{j_1}-P_{j_1}\big\vert>\epsilon/3\big\}\leq 2p\exp\big(-C_{\epsilon}n\big)$. By condition (C1), we know that $0< P_{\min}^{(1)}\leq P_j\leq P_{\max}^{(1)}<1$ for every $ 1\leq j\leq p$. Then, we can obtain that $ P\Big\{\wt{P}_{j_1}\in\big(P_{\min}^{(1)}-\epsilon/3, P_{\max}^{(1)}+\epsilon/3\big)\Big\}\geq 1-2p\exp\big(-C_{\epsilon}n\big)$ for every $ 1\leq j_1\leq p $. Note that $ \dot{\Phi}^{-1}(\cdot)=\big\{\phi\big(\Phi^{-1}(\cdot)\big)\big\}^{-1} $ is a continous function and is independent of $j_1$. Then, re-write $M$  as $M_{\epsilon}=\max\Big\{\sup_{P_{\min}^{(1)}-\epsilon/3<\tau< P_{\max}^{(1)}+\epsilon/3}\big\vert\phi\big(\Phi^{-1}(\tau)\big)\big\vert^{-1},1\Big\}$. Therefore, $\mQ_1^{b_2}=P\big\{\max_{j_1}\sup_{0<\eta<1}\big\vert\dot{\Phi}^{-1}$  $\big(\wt{P}_{j_1}\big)\big\vert>M_{\epsilon}\big\}\leq 2p\exp\big(-C_{\epsilon}n\big)$. Combining the results of $\mQ_1^{b_1}$ and $\mQ_1^{b_2} $, we have proved $ \mQ_1^b\leq \mQ_1^{b_1}+\mQ_1^{b_2}\leq  4p\exp\big\{-C_{\epsilon}^an\big\}$ with $C_{\epsilon}^a=2\epsilon^2/\big(9M_{\epsilon}^2+4M_{\epsilon}\epsilon\big)$. Identically, we can obtain that $ \mQ_1^c\leq 4p\exp\big\{-C_{\epsilon}^an\big\} $. Furthermore, combining the above results, we complete the proof of $\mQ_1=P\big\{\max_{j_1,j_2}\big\Vert \wh{\theta}_{j_1j_2}-\theta_{j_1j_2}\big\Vert>\epsilon\big\}\leq 4p^2\exp\big(-C_{\epsilon}^an\big)$ as long as $p>2$.

\sc{Step 2.}\normalfont$\ $Next, we study $\mQ_2= P\big\{\max_{j_1,j_2}\sup_{0\leq\eta\leq1}\big\Vert\ddot{g}\big(\wt{\theta}_{j_1j_2}\big)\big\Vert>M_{\epsilon}^g\big\}$. Recall that $\ddot{g}\big(\theta_{j_1j_2}\big)=\partial^2 g\big(\theta_{j_1j_2}\big)/\partial\theta_{j_1j_2}\partial\theta_{j_1j_2}^{\top}\in\mR^{3\times 3}$. Write $\ddot{g}\big(\theta_{j_1j_2}\big)=\big(h^{k_1k_2}\big(\theta_{j_1j_2}\big)\big)\in\mR^{3\times 3}$, where $h^{k_1k_2}\big(\theta_{j_1j_2}\big)=\partial^2 g\big(\theta_{j_1j_2}\big)/\partial\theta_{j_1j_2}^{k_1}\partial\theta_{j_1j_2}^{k_2\top}$. Note that $h^{k_1k_2}(\cdot)$ is a continuous function and is independent of $j_1$ and $j_2$ for every $ 1\leq k_1,k_2\leq 3 $. In the meanwhile, we know that $\max_{j_1,j_2}\big\Vert \wt{\theta}_{j_1j_2}-\theta_{j_1j_2}\big\Vert\leq \max_{j_1,j_2}\big\Vert \wh{\theta}_{j_1j_2}-\theta_{j_1j_2}\big\Vert$. By \sc{Step 1,}\normalfont$\ $we have $ P\big\{\max_{j_1,j_2}\big\Vert \wt{\theta}_{j_1j_2}-\theta_{j_1j_2}\big\Vert>\epsilon\big\}\leq P\big\{\max_{j_1,j_2}\big\Vert \wh{\theta}_{j_1j_2}-\theta_{j_1j_2}\big\Vert>\epsilon\big\}\leq 4p^2\exp\big(-C_{\epsilon}^an\big)$ for any given $ \epsilon>0$. By condition (C2), we know that $0< P_{\min}^{(2)}\leq P_{j_1j_2}\leq P_{\max}^{(2)}<1$ for every $ 1\leq j_1,j_2\leq p$. In the meanwhile, we have $ \vert c_j\vert\leq c_{\max}$ for every $1\leq j\leq p$, where $c_{\max}=\max\big\{\vert\Phi^{-1}\big(P_{\min}^{(1)}\big)\vert,\vert\Phi^{-1}\big(P_{\max}^{(1)}\big)\vert\big\}$.
Write $\theta_{\min}=P_{\min}^{(2)}$ and $\theta_{\max}=\big\{2c_{\max}^2+\big(P_{\max}^{(2)}\big)^2\big\}^{1/2}$. Then, $\theta_{\min}\leq\big\Vert\theta_{j_1j_2}\big\Vert\leq \theta_{\max}$ for every $ 1\leq j_1,j_2\leq p$. Subsequently, we can obtain that 
\begin{equation}\label{equ:theta}
	P\Big\{\big\Vert\wt{\theta}_{j_1j_2}\big\Vert\in\big(\theta_{\min}-\epsilon,\theta_{\max}+\epsilon\big)\Big\}\geq 1-4p^2\exp\big(-C_{\epsilon}^an\big).
\end{equation}
Define $ M_{\epsilon}^h=\max\Big\{\max_{k_1,k_2}\sup_{\theta_{\min}-\epsilon<\Vert\tau\Vert< \theta_{\max}+\epsilon}\big\vert h^{k_1k_2}(\tau)\big\vert,1\Big\}$. We then have $P\big\{$ $\max_{j_1,j_2}\sup_{0\leq\eta\leq1}\big\vert h^{k_1k_2}\big(\wt{\theta}_{j_1j_2}\big)\big\vert>M_{\epsilon}^h\big\}\leq 4p^2\exp\big(-C_{\epsilon}^an\big)$. Note that $\ddot{g}\big(\wt{\theta}_{j_1j_2}\big)$ is a symmetric matrix. Thus, by the proof of Lemma 1 in \cite{2009Wang}, we have $\big\Vert\ddot{g}\big(\wt{\theta}_{j_1j_2}\big)\big\Vert\leq 3\max_{k_1,k_2}\big\vert h^{k_1k_2}\big(\wt{\theta}_{j_1j_2}\big)\big\vert$. Let $ M_{\epsilon}^g=3M_{\epsilon}^h$. Then, $ \mQ_2\leq P\big\{\max_{j_1,j_2}$ $\sup_{0\leq\eta\leq1}\max_{k_1,k_2}\big\vert h^{k_1k_2}\big(\wt{\theta}_{j_1j_2}\big)\big\vert>M_{\epsilon}^h\big\}\leq\sum_{k_1,k_2}P\big\{\max_{j_1,j_2}\sup_{0\leq\eta\leq1}\big\vert h^{k_1k_2}\big(\wt{\theta}_{j_1j_2}\big)\big\vert>M_{\epsilon}^h\big\} \leq 36p^2\exp\big(-C_{\epsilon}^an\big)$. This completes the proof of $ \mQ_2\leq 36p^2\exp\big(-C_{\epsilon}^an\big)$.

\sc{Step 3.}\normalfont$\ $Lastly, we prove the theorem conclusion. We focus on $P^*=P\big\{\max_{1\leq j_1\neq j_2\leq p}$ $\big\vert \wh{\sigma}_{j_1j_2}-\sigma_{j_1j_2}-\dot{g}\big(\theta_{j_1j_2}\big)^{\top}\big(\wh{\theta}_{j_1j_2}-\theta_{j_1j_2}\big)\big\vert>\epsilon\big\}$ for any given $\epsilon>0$. Recall that $\wh{\sigma}_{j_1j_2}=g\big(\wh{\theta}_{j_1j_2}\big)$. Then, we can conduct the Taylor's expansion about $\wh{\sigma}_{j_1j_2}=g\big(\wh{\theta}_{j_1j_2}\big)$ for $\wh{\theta}_{j_1j_2}$ at $\theta_{j_1j_2}$ as
\begin{equation}\label{equ:taylor}
	\wh{\sigma}_{j_1j_2}-\sigma_{j_1j_2}=\dot{g}\big(\theta_{j_1j_2}\big)^{\top}\big(\wh{\theta}_{j_1j_2}-\theta_{j_1j_2}\big)+\frac{1}{2}\big(\wh{\theta}_{j_1j_2}-\theta_{j_1j_2}\big)^{\top}\ddot{g}\big(\wt{\theta}_{j_1j_2}\big)\big(\wh{\theta}_{j_1j_2}-\theta_{j_1j_2}\big),
\end{equation}
where $\wt{\theta}_{j_1j_2}=\eta_{j_1j_2}\wh{\theta}_{j_1j_2}+(1-\eta_{j_1j_2})\theta_{j_1j_2}$ for some $\eta_{j_1j_2}\in(0,1)$. We then have $ \max_{j_1,j_2}$ $\big\vert \wh{\sigma}_{j_1j_2}-\sigma_{j_1j_2}-\dot{g}\big(\theta_{j_1j_2}\big)^{\top}\big(\wh{\theta}_{j_1j_2}-\theta_{j_1j_2}\big)\big\vert=\max_{j_1,j_2}\big\vert2^{-1}\big(\wh{\theta}_{j_1j_2}-\theta_{j_1j_2}\big)^{\top}\ddot{g}\big(\wt{\theta}_{j_1j_2}\big)\big(\wh{\theta}_{j_1j_2}-\theta_{j_1j_2}\big)\big\vert\leq 2^{-1}\max_{j_1,j_2}\sup_{0<\eta<1}\big\Vert\ddot{g}\big(\wt{\theta}_{j_1j_2}\big)\big\Vert\max_{j_1,j_2}\big\Vert\wh{\theta}_{j_1j_2}-\theta_{j_1j_2}\big\Vert^2$. In this regard, we have $P^*\leq W_1+W_2$, where $W_1=P\big\{\max_{1\leq j_1\neq j_2\leq p}\big\Vert \wh{\theta}_{j_1j_2}-\theta_{j_1j_2}\big\Vert^2>2\epsilon/M_{\epsilon}^g\big\}$ and $W_2=P\big\{\max_{1\leq j_1\neq j_2\leq p}$  $\sup_{0<\eta<1}\big\Vert\ddot{g}\big(\wt{\theta}_{j_1j_2}\big)\big\Vert>M_{\epsilon}^g\big\}$. Subsequently, it suffices to upper bound $W_1$ and $W_2$ separately. By \sc{Step 1,}\normalfont$\ $we have $P\big\{\max_{1\leq j_1\neq j_2\leq p}\big\Vert \wh{\theta}_{j_1j_2}-\theta_{j_1j_2}\big\Vert>\epsilon\big\}\leq 4p^2\exp\big(-C_{\epsilon}^an\big)$. Then, we can obtain that $W_1\leq 4p^2\exp\big(-C_{\epsilon}^bn\big)$, where $C_{\epsilon}^b=4\epsilon\big/\big(9M_{\epsilon}^2M_{\epsilon}^g+4M_{\epsilon}\sqrt{2M_{\epsilon}^g\epsilon}\big)$. By \sc{Step 2,}\normalfont$\ $we have $W_2\leq \mQ_2\leq 36p^2\exp\big(-C_{\epsilon}^an\big)$, where $C_{\epsilon}^a=2\epsilon^2/\big(9M_{\epsilon}^2+4M_{\epsilon}\epsilon\big)$. Combining the above results, $P^*$ is upper bounded by $4p^2\exp\big\{-C_{\epsilon}^{b}n\big\}+36p^2\exp\big(-C_{\epsilon}^an\big)\leq 40p^2\exp\big\{-C_{\epsilon}^*n\big\}$, where $C_{\epsilon}^*=\min\big(C_{\epsilon}^a,C_{\epsilon}^b\big)$. This completes the whole proof of Theorem 1.

\subsection*{\textbf{Appendix A.2. Proof of Theorem 2}}

Define $\mD^*(B,\Sigma_d)=p^{-1}\tr\big(B^{\top}Q_{\Sigma_d}B\big)$, where $H_{\Sigma_d}=\Sigma_d(\Sigma_d^{\top}\Sigma_d)^{-1}\Sigma_d^{\top}=\Sigma_d\Sigma_d^{\top}$ is the projection matrix onto the linear subspace of $ \mS(\Sigma_d) $ and $ Q_{\Sigma_d}=I_p-H_{\Sigma_d} $ is the projection matrix onto the orthogonal linear subspace of $ \mS(\Sigma_d)$. Then, the desired conclusion can be proved in a total of six steps. In the 1st step, we prove that $\big\vert p^{-1}\sum_{k=1}^d\lambda_k\big(\Sigma\big)-\sum_{k=1}^d\lambda_k\big(\Sigma_B\big)\big\vert\to 0$ as $ p\to\infty$. In the 2nd step, we show that $ \mD^*\big(B,\Sigma_d\big)\to 0 $ as $ p\to\infty $. In the 3rd step, we show that $ \mD\big(B,\Sigma_d\big)\to 0 $ as $ p\to\infty $.
In the 4th step, we prove that $\big\Vert\wh{\Sigma}-\Sigma\big\Vert=o_p(1)$ as $ n\to\infty$. In the 5th step, we prove that $ \mD^*\big(B,\wh{\Sigma}_d\big)=o_p(1) $ as $ n\to\infty $. In the last step, we show that $ \mD\big(B,\wh{\Sigma}_d\big)\xrightarrow{p}0$ as $ n\to\infty $. The details are given below.

\sc{Step 1.}\normalfont$\ $We start with $\big\vert p^{-1}\sum_{k=1}^d\lambda_k\big(\Sigma\big)-\sum_{k=1}^d\lambda_k\big(\Sigma_B\big)\big\vert$. By definition, we can obtain that $ p^{-1}\sum_{k=1}^d\lambda_k\big(\Sigma\big)=\sup_{\mB^{\top}\mB=I_d}$ $p^{-1}\tr\big(\mB^{\top}\Sigma \mB\big)$. Note that $\Sigma=BB^{\top}+D$. We then have 
\begin{equation}\label{bb}
	\frac{1}{p}\sum_{k=1}^d\lambda_k\big(\Sigma\big)=\sup_{\mB^{\top}\mB=I_d}\bigg[\frac{1}{p}\tr\Big\{\mB^{\top}\big(BB^{\top}\big)\mB\Big\}+\frac{1}{p}\tr\big(\mB^{\top}D\mB\big)\bigg].
\end{equation}
Then, we have $\mQ_1^a-\mQ_1^b\leq p^{-1}\sum_{k=1}^d\lambda_k\big(\Sigma\big)\leq \mQ_1^a+\mQ_1^b$, where $\mQ_1^a=\sup_{\mB^{\top}\mB=I_d}p^{-1}$ $\tr\big\{\mB^{\top}\big(BB^{\top}\big)$ $\mB\big\}$ and $\mQ_1^b=\sup_{\mB^{\top}\mB=I_d}p^{-1}\tr\big(\mB^{\top}D\mB\big)$. Note that $\mQ_1^a=\sup_{\mB^{\top}\mB=I_d}$ $\tr\big\{\mB^{\top}\big(\Sigma_B\big)\mB\big\}=\sum_{k=1}^d\lambda_k\big(\Sigma_B\big)$. Therefore, it suffices to show that $\mQ_1^b\to 0$ as $ p\to\infty$. By definition, we have $ \mQ_1^b=\sup_{\mB^{\top}\mB=I_d}p^{-1}\tr\big(D\mB\mB^{\top}\big)$. Note that $ D $ is a positive definite matrix and $\mB\mB^{\top}$ is a semi-positive definite matrix. Then by Lemma 2 in \cite{wang2012} and condition  (C3), we can obtain that $\mQ_1^b\leq\sup_{\mB^{\top}\mB=I_d}\lambda_{\max}\big(D\big)\tr\big(\mB\mB^{\top}\big)/p=\tau_{\max}^2d/p$. Thus, we have $\mQ_1^b\to 0$ as $ p\to\infty$. Furthermore, this completes the proof of $\big\vert p^{-1}\sum_{k=1}^d\lambda_k\big(\Sigma\big)-\sum_{k=1}^d\lambda_k\big(\Sigma_B\big)\big\vert\to 0$ as $ p\to\infty$.

\sc{Step 2.}\normalfont$\ $We next prove that $\mD^*\big(B,\Sigma_d\big)\to 0 $ as $ p\to\infty$. By definition, we have $ \mD^*\big(B,\Sigma_d\big)=p^{-1}\tr\big(B^{\top}Q_{\Sigma_d}B\big)=p^{-1}\tr\big(B^{\top}B\big)-p^{-1}\tr\big(B^{\top}H_{\Sigma_d}B\big)$. Note that $ \Sigma_B=B^{\top}B/p=\diag\big\{\lambda_1(\Sigma_B),\ldots,\lambda_d(\Sigma_B)\big\}$. We then have $ p^{-1}\tr\big(B^{\top}B\big)=\sum_{k=1}^d\lambda_k\big(\Sigma_B\big)$. Recall that $ \Sigma=BB^{\top}+D$. Then, we can obtain that $p^{-1}\tr\big(B^{\top}H_{\Sigma_d}B\big)=p^{-1}\tr\big(H_{\Sigma_d}BB^{\top}\big)$ $=p^{-1}\tr\big(H_{\Sigma_d}\Sigma\big)-p^{-1}\tr\big(H_{\Sigma_d}D\big)$. By definition, we have $ p^{-1}\tr\big(H_{\Sigma_d}\Sigma\big)=p^{-1}\sum_{k=1}^d\lambda_k\big(\Sigma\big)$. Thus, we can obtain that $ \mD^*\big(B,\Sigma_d\big)\leq \mQ_2^a+\mQ_2^b$, where $ \mQ_2^a=\big\vert \sum_{k=1}^d\lambda_k\big(\Sigma_B\big)-p^{-1}\sum_{k=1}^d\lambda_k\big(\Sigma\big)\big\vert $ and $ \mQ_2^b=\big\vert p^{-1}\tr\big(H_{\Sigma_d}D\big)\big\vert $. By \sc{Step 1,}\normalfont$\ $we know that $ \mQ_2^a\to 0 $ as $ p\to\infty $. Therefore, we should focus on $ \mQ_2^b $. Note that $ D$ is a positive definite matrix and $H_{\Sigma_d}$ is a semi-positive definite matrix. By Lemma 2 in \cite{wang2012} and condition (C3), we can obtain that $ \mQ_2^b\leq  p^{-1}\lambda_{\max}\big(D\big)\tr\big(H_{\Sigma_d}\big)=\tau_{\max}^2d/p$. Combining the results of $ \mQ_2^a $ and $ \mQ_2^b $, we have $ \mD^*\big(B,\Sigma_d\big)\to 0 $ as $ p\to\infty$.

\sc{Step 3.}\normalfont$\ $Next, we study $\mD\big(B,\Sigma_d\big)$. By \sc{Step 2,}\normalfont$\ $we know that $ \mD^*\big(B,\Sigma_d\big)\to 0 $ as $ p\to\infty$. Since both $\mD\big(B,\Sigma_d\big)$ and $ \mD^*\big(B,\Sigma_d\big)$ are discrepancy measures between $ B $ and $ \Sigma_d $, we shall study the relationship between $\mD\big(B,\Sigma_d\big)$ and $ \mD^*\big(B,\Sigma_d\big)$. To this end, we first consider $ \mT=p^{-1}\tr\big(B^{\top}Q_{\Sigma_d}B\Sigma_B^{-1}\big)$. Note that $ B^{\top}Q_{\Sigma_d}B $ is a semi-positive definite matrix and $ \Sigma_B=B^{\top}B/p $ is a positive definite matrix. By Lemma 2 in \cite{wang2012} and Section 2.3, we can obtain that $ \mT\leq\lambda_{\min}^{-1}\big(\Sigma_B\big)p^{-1}\tr\big(B^{\top}Q_{\Sigma_d}B\big)=\lambda_d^{-1}\big(\Sigma_B\big)\mD^*\big(B,\Sigma_d\big)\to 0 $ as $ p\to\infty$. In the meanwhile, direct computation leads to $ \mT=\big\{B^{\top}Q_{\Sigma_d}B(B^{\top}B)^{-1}\big\}=\tr\big\{(I_p-H_{\Sigma_d})H_B\big\}=\tr\big(H_B\big)-\tr\big(H_{\Sigma_d}H_B\big)=d-\tr\big(H_BH_{\Sigma_d}\big)$. Next, we compute $ \mD\big(B,\Sigma_d\big) $. By definition, we have $ \mD\big(B,\Sigma_d\big)=\tr\big(H_B-H_{\Sigma_d}\big)^2=\tr\big(H_B^2\big)+\tr\big(H_{\Sigma_d}^2\big)-2\tr\big(H_BH_{\Sigma_d}\big)=2\big\{d-\tr\big(H_BH_{\Sigma_d}\big)\big\}=2\mT\to 0$ as $ p\to\infty$. This completes the proof of the 1st conclusion of Theorem 2.

\sc{Step 4.}\normalfont$\ $To establish the 2nd conclusion of Theorem 2, we need to show that $\big\Vert\wh{\Sigma}-\Sigma\big\Vert=o_p(1)$ as $ n\to\infty$. In this regard, it suffices to study $P^*=P\big\{\big\Vert\wh{\Sigma}-\Sigma\big\Vert>2\epsilon\big\}$ for any given $ \epsilon>0 $. By the proof of Lemma 1 in \cite{2009Wang}, we can obtain that $\big\Vert\wh{\Sigma}-\Sigma\big\Vert\leq p\max_{j_1,j_2}\big\vert\wh{\sigma}_{j_1j_2}-\sigma_{j_1j_2}\big\vert$. Recall that $\wh{\sigma}_{jj}=\sigma_{jj}=1$ in Section 2.1. We then have $ P^*\leq P\big\{\max_{1\leq j_1\neq j_2\leq p}\big\vert\wh{\sigma}_{j_1j_2}-\sigma_{j_1j_2}\big\vert>2\epsilon/p\big\}$. Note that $ \max_{1\leq j_1\neq j_2\leq p}\big\vert\wh{\sigma}_{j_1j_2}-\sigma_{j_1j_2}\big\vert\leq \max_{1\leq j_1\neq j_2\leq p}\big\vert \wh{\sigma}_{j_1j_2}-\sigma_{j_1j_2}-\dot{g}\big(\theta_{j_1j_2}\big)^{\top}\big(\wh{\theta}_{j_1j_2}-\theta_{j_1j_2}\big)\big\vert+ \max_{j_1,j_2}\big\vert\dot{g}\big(\theta_{j_1j_2}\big)^{\top}\big(\wh{\theta}_{j_1j_2}-\theta_{j_1j_2}\big)\big\vert$. Therefore, we further have $ P^*\leq \mQ_4^a+\mQ_4^b$, where $ \mQ_4^a=P\big\{\max_{1\leq j_1\neq j_2\leq p}\big\vert \wh{\sigma}_{j_1j_2}-\sigma_{j_1j_2}-\dot{g}\big(\theta_{j_1j_2}\big)^{\top}\big(\wh{\theta}_{j_1j_2}-\theta_{j_1j_2}\big)\big\vert>\epsilon/p\big\} $ and $ \mQ_4^b=P\big\{\max_{j_1,j_2}\big\vert \dot{g}\big(\theta_{j_1j_2}\big)^{\top}$  $\big(\wh{\theta}_{j_1j_2}-\theta_{j_1j_2}\big)\big\vert>\epsilon/p\big\}$. To upper bound $ P^* $, it suffices to bound $ \mQ_4^a $ and $ \mQ_4^b $ separately. By the proof of theorem 1, we know that there exits some sufficiently large but fixed constants $ M_{\epsilon}>0$ and $ M_{\epsilon}^g>0$ such that $\mQ_4^a\leq 40p^2\exp\big\{-C_{\epsilon}^*(p)np^{-2}\big\}\leq 40\exp\big\{2\log(p)-C_{\epsilon}^*(p)np^{-2}\big\}$, where $C_{\epsilon}^*(p)=\min\Big\{2\epsilon^2/\big(9M_{\epsilon}^2+4M_{\epsilon}\epsilon/p\big),4\epsilon\big/\big(9M_{\epsilon}^2M_{\epsilon}^g+4M_{\epsilon}\sqrt{2M_{\epsilon}^g\epsilon/p}\big)\Big\}$. By condition (C4), we have $ p\leq \rho n^{\xi}$. Note that $ C_{\epsilon}^*(p)\to C_{\epsilon}^{*a}=\min\Big\{2\epsilon^2/\big(9M_{\epsilon}^2\big),$ $4\epsilon\big/\big(9M_{\epsilon}^2M_{\epsilon}^g\big)\Big\}$ as $p\to\infty$. Then, $ \mQ_4^a$ can be further bounded by $ \mQ_4^a\leq 40\rho^2\exp\Big[ \log(n)\big\{2\xi-C_{\epsilon}^{*a}\rho^{-2}n^{1-2\xi}/\log(n)\big\}\Big]$. By condition (C4), we have $ \xi<1/2$. Thus, the upper bound of $ \mQ_4^a$ converges to 0 as $n\to\infty$. Next, we focus on $ \mQ_4^b $.

Direct computation leads to $ \max_{j_1,j_2}\big\vert \dot{g}\big(\theta_{j_1j_2}\big)^{\top}\big(\wh{\theta}_{j_1j_2}-\theta_{j_1j_2}\big)\big\vert\leq \max_{j_1,j_2}\big\Vert \dot{g}\big(\theta_{j_1j_2}\big)\big\Vert$  $\max_{j_1,j_2}\big\Vert\wh{\theta}_{j_1j_2}-\theta_{j_1j_2}\big\Vert $. Recall that $\theta_{j_1j_2}=(c_{j_1},c_{j_2},P_{j_1j_2})^{\top}$ and $\dot{g}\big(\theta_{j_1j_2}\big)=\partial g\big(\theta_{j_1j_2}\big)/\partial\theta_{j_1j_2}$ $\in\mR^3$. Write $ \dot{g}\big(\theta_{j_1j_2}\big)=\big(\dot{g}_1\big(\theta_{j_1j_2}\big),\dot{g}_2\big(\theta_{j_1j_2}\big),\dot{g}_3\big(\theta_{j_1j_2}\big)\big)^{\top}$. Note that $ \dot{g}_k\big(\theta_{j_1j_2}\big) $ is a continuous function and is independent of $ j_1 $ and $ j_2$ for every $ 1\leq k\leq 3 $. In the meanwhile, by the proof of \sc{Step 2}\normalfont$\ $of Theorem 1, we have $ \theta_{\min}\leq\big\Vert\theta_{j_1j_2}\big\Vert\leq \theta_{\max}$ for every $ 1\leq j_1,j_2\leq p$. Therefore, there exists an sufficiently large but fixed constant $ M_{\epsilon}^{g_1}>0$ such that $\max_{j_1,j_2}\big\Vert \dot{g}\big(\theta_{j_1j_2}\big)\big\Vert\leq M_{\epsilon}^{g_1}$. We then have $ \mQ_4^b\leq P\big\{\max_{j_1,j_2}\big\Vert\wh{\theta}_{j_1j_2}-\theta_{j_1j_2}\big\Vert>\epsilon/\big(pM_{\epsilon}^{g_1}\big)\big\} $. By \sc{Step 2}\normalfont$\ $of Theorem 1, we can directly have $\mQ_4^b\leq 4p^2\exp\big\{-C_{\epsilon}^b(p)np^{-2}\big\}$, where $ C_{\epsilon}^b(p)=2\epsilon^2/\big\{9\big(M_{\epsilon}M_{\epsilon}^{g_1}\big)^2+4M_{\epsilon}M_{\epsilon}^{g_1}\epsilon/p\big\} $. Similar to $\mQ_4^a$, we can obtain that $\mQ_4^b\leq  4\rho^2\exp\Big[\log(n)\big\{2\xi-C_{\epsilon}^{*b}\rho^{-2}n^{1-2\xi}/\log(n)\big\}\Big]$, where $ C_{\epsilon}^{*b}=2\epsilon^2/\big\{9\big(M_{\epsilon}M_{\epsilon}^{g_1}\big)^2\big\} $. Thus, $\mQ_4^b\to 0 $ as $n\to\infty$. Combining the results of $ \mQ_4^a$ and $ \mQ_4^b $, we then have $ P^*=P\big\{\big\Vert\wh{\Sigma}-\Sigma\big\Vert>2\epsilon\big\}\leq \mQ_4^a+\mQ_4^b\to 0$ as $ n\to\infty$. Then, we can obtain that $\big\Vert\wh{\Sigma}-\Sigma\big\Vert=o_p(1) $ as $ n\to\infty$.

\sc{Step 5.}\normalfont$\ $Next, we consider $\mD^*(B,\wh{\Sigma}_d)=p^{-1}\tr\big(B^{\top}Q_{\wh{\Sigma}_d}B\big)$, where $Q_{\wh{\Sigma}_d}=I_p-H_{\wh{\Sigma}_d}$, $ H_{\wh{\Sigma}_d}=\wh{\Sigma}_d\big(\wh{\Sigma}_d^{\top}\wh{\Sigma}_d\big)^{-1}\wh{\Sigma}_d^{\top}$ and $\wh{\Sigma}_d=(\wh{v}_1,\ldots,\wh{v}_d)\in\mR^{p\times d}$. Then, we have $ \wh{\Sigma}_d^{\top}\wh{\Sigma}_d=I_d$ and $H_{\wh{\Sigma}_d}=\wh{\Sigma}_d\wh{\Sigma}_d^{\top}$. Obviously, we have $ \mD^*(B,\wh{\Sigma}_d)\geq 0 $. Recall that $ p^{-1}\tr\big(B^{\top}B\big)=\sum_{k=1}^d\lambda_k\big(\Sigma_B\big)$. Then direct computation leads to $\mD^*(B,\wh{\Sigma}_d)=p^{-1}\tr\big(B^{\top}B\big)-p^{-1}\tr\big(B^{\top}$  $H_{\wh{\Sigma}_d}B\big)\leq \mQ_5^a+\mQ_5^b$, where $\mQ_5^a=\big\vert\sum_{k=1}^d\lambda_k\big(\Sigma_B\big)-p^{-1}\sum_{k=1}^d\lambda_k\big(\Sigma\big)\big\vert$ and $\mQ_5^b=\big\vert p^{-1}\sum_{k=1}^d$  $\lambda_k\big(\Sigma\big)-p^{-1}\tr\big(H_{\wh{\Sigma}_d}BB^{\top}\big)\big\vert$. To prove $\mD^*(B,\wh{\Sigma}_d)=o_p(1)$, it suffices to upper bound $\mQ_5^a$ and $\mQ_5^b$ separately. By \sc{Step 1,}\normalfont$\ $we have proved $\mQ_5^a\to 0 $ as $ p\to\infty $. To study $\mQ_5^b$, we first compute $p^{-1}\tr\big(H_{\wh{\Sigma}_d}$  $BB^{\top}\big)$. Recall that $ \Sigma=BB^{\top}+D $. Then, we have $ p^{-1}\tr\big(H_{\wh{\Sigma}_d}BB^{\top}\big)=p^{-1}\tr\big(H_{\wh{\Sigma}_d}\Sigma\big)$  $-p^{-1}\tr\big(H_{\wh{\Sigma}_d}D\big)=p^{-1}\tr\big(H_{\wh{\Sigma}_d}\wh{\Sigma}\big)+p^{-1}\tr\big\{H_{\wh{\Sigma}_d}\big(\Sigma-\wh{\Sigma}\big)\big\}-p^{-1}\tr\big(H_{\wh{\Sigma}_d}D\big)$. Note that $p^{-1}\tr\big(H_{\wh{\Sigma}_d}\wh{\Sigma}\big)=p^{-1}\sum_{k=1}^d\lambda_k\big(\wh{\Sigma}\big)$. Thus, we can obtain that $\mQ_5^b\leq W_1+W_2+W_3$, where $ W_1=p^{-1}\big\vert\sum_{k=1}^d\lambda_k\big(\wh{\Sigma}\big)-\sum_{k=1}^d\lambda_k\big(\Sigma\big)\big\vert$, $ W_2=p^{-1}\big\vert\tr\big\{H_{\wh{\Sigma}_d}\big(\wh{\Sigma}-\Sigma\big)\big\}\big\vert $ and $ W_3=p^{-1}\big\vert\tr\big(H_{\wh{\Sigma}_d}D\big)\big\vert $. To upper bound $\mQ_5^b$, it sffices to upper bound $ W_1 $, $W_2$ and $W_3$ separately. 

\sc{Step 5.1.}\normalfont$\ $We start with $W_1=p^{-1}\big\vert \sum_{k=1}^d\lambda_k\big(\wh{\Sigma}\big)-\sum_{k=1}^d\lambda_k\big(\Sigma\big)\big\vert$. By definition, we can obtain that 
\begin{equation}\label{equ:chai}
	\frac{1}{p}\sum_{k=1}^d\lambda_k\big(\wh{\Sigma}\big)=\sup_{\mB^{\top}\mB=I_d}\frac{1}{p}\tr\Big(\mB^{\top}\wh{\Sigma} \mB\Big)=\sup_{\mB^{\top}\mB=I_d}\bigg[\frac{1}{p}\tr\big(\mB^{\top}\Sigma \mB\big)+\frac{1}{p}\tr\big\{\mB^{\top}(\wh{\Sigma}-\Sigma)\mB\big\}\bigg]
\end{equation}
Then, we have $\mO_5^a-\mO_5^b\leq p^{-1}\sum_{k=1}^d\lambda_k\big(\wh{\Sigma}\big)\leq \mO_5^a+\mO_5^b$, where $\mO_5^a=\sup_{\mB^{\top}\mB=I_d}p^{-1}\tr\big\{\mB^{\top}\Sigma$  $ \mB\big\}=p^{-1}\sum_{k=1}^d\lambda_k\big(\Sigma\big)$ and $\mO_5^b=\sup_{\mB^{\top}\mB=I_d}p^{-1}\tr\big(\mB^{\top}\big(\wh{\Sigma}-\Sigma\big)\mB\big)$. Therefore, it suffices to show $\mO_5^b=o_p(1) $ as $ n\to\infty$. By definition, we have $ \mO_5^b=\sup_{\mB^{\top}\mB=I_d}p^{-1}\tr\big\{\big(\wh{\Sigma}-\Sigma\big)\mB\mB^{\top}\big\}$. Note that $\wh{\Sigma}-\Sigma $ is a symmetric matrix. Then by Lemma 2 in \cite{wang2012}, we can obtain that $\mO_5^b\leq\sup_{\mB^{\top}\mB=I_d}\big\Vert\wh{\Sigma}-\Sigma\big\Vert\tr\big(\mB\mB^{\top}\big)/p=d\big\Vert\wh{\Sigma}-\Sigma\big\Vert/p$. By \sc{Step 4}\normalfont$\ $, we know that $\big\Vert\wh{\Sigma}-\Sigma\big\Vert=o_p(1)$ as $ n\to\infty $. Then, we have $\mO_5^b=o_p(1)$. Therefore, we have $p^{-1}\big\vert \sum_{k=1}^d\lambda_k\big(\wh{\Sigma}\big)-\sum_{k=1}^d\lambda_k\big(\Sigma\big)\big\vert=o_p(1)$ as $n\to\infty$.

\sc{Step 5.2.}\normalfont$\ $Next, we study $ W_2 $ and $ W_3$. We start with $ W_2=p^{-1}\big\vert\tr\big\{H_{\wh{\Sigma}_d}\big(\wh{\Sigma}-\Sigma\big)\big\}\big\vert$. Note that $H_{\wh{\Sigma}_d}$ is a projecting matrix of rank $d$ and then we have $\tr\big(H_{\wh{\Sigma}_d}\big)=d$. By Lemma 2 in \cite{wang2012} and \sc{Step 5.1,}\normalfont$\ $we can obtain that $W_2\leq \big\Vert\wh{\Sigma}-\Sigma\big\Vert\tr\big(H_{\wh{\Sigma}_d}\big)/p=d\big\Vert\wh{\Sigma}-\Sigma\big\Vert/p=o_p(1)$. Thus, we have $W_2=o_p(1)$ as $ n\to\infty $. Similar to \sc{Step 5.1,}\normalfont$\ $we can obtain that $ W_3\leq \lambda_{\max}(D)\tr\big(H_{\wh{\Sigma}_d}\big)/p=\tau_{\max}^2d/p$. Recall that $\mQ_5^a \to0$ as $ p\to\infty $ and condition (C4). We then have $\mD^*(B,\wh{\Sigma}_d)\leq Q_5^a+Q_5^b\leq Q_5^a+W_1+W_2+W_3=o_p(1)$ as $ n\to\infty $. This completes the proof of $ \mD^*(B,\wh{\Sigma}_d)=o_p(1)$ as $ n\to\infty $.

\sc{Step 6.}\normalfont$\ $Lastly, we prove that $ \mD\big(B,\wh{\Sigma}_d\big)\xrightarrow{p}0$ as $ n\to\infty $. By \sc{Step 5,}\normalfont$\ $we know that $ \mD^*(B,\wh{\Sigma}_d)=o_p(1)$ as $ n\to\infty $. Similar to \sc{Step 3,}\normalfont$\ $we study the relationship between $ \mD\big(B,\wh{\Sigma}_d\big)$ and $\mD^*\big(B,\wh{\Sigma}_d\big)$. Here we consider $ \mT^*=p^{-1}\tr\big(B^{\top}Q_{\wh{\Sigma}_d}B\Sigma_B^{-1}\big)$. It can be verified that $ \mT^*=o_p(1)$ as $ n\to\infty $ and $ \mT^*=d-\tr\big(H_{\wh{\Sigma}_d}H_B\big)$. Similar to \sc{Step 3,}\normalfont$\ $direct computation leads to $ \mD\big(B,\wh{\Sigma}_d\big)=2\big\{d-\tr\big(H_BH_{\wh{\Sigma}_d}\big)\big\}=2\mT^*$. Then, we have $\mD\big(B,\wh{\Sigma}_d\big)\xrightarrow{p}0$ as $ n\to\infty$. This completes the proof of the 2nd conclusion of Theorem 2.

\subsection*{\textbf{Appendix A.3. Proof of Theorem 3}}

Recall that $\Sigma=BB^{\top}+D$ and $D=\diag\big\{\tau_1^2,\ldots,\tau_p^2\big\}$. By Section 2.4, we know that $\wh{D}=\diag\Big\{Q_{\wh{\Sigma}_d}\wh{\Sigma}Q_{\wh{\Sigma}_d}^{\top}\Big\}$, where $Q_{\wh{\Sigma}_d}=Q_{\wh{\Sigma}_d}^{\top}=I_p-H_{\wh{\Sigma}_d}$ and $ H_{\wh{\Sigma}_d}=\wh{\Sigma}_d\wh{\Sigma}_d^{\top}$. Let $A=(a_{ij})\in\mR^{M\times N}$ be an arbitrary matrix. Define its absolute value as $\vert A\vert=\big(\vert a_{ij}\vert\big)\in\mR^{M\times N}$ \citep{van2020}. We then have $ p^{-1}\sum_{j=1}^{p}\big\vert\wh{\tau}_j^2-\tau_j^2\big\vert=p^{-1}\tr\Big\{\big\vert\wh{D}-D\big\vert\Big\}=p^{-1}\tr\Big\{\Big\vert Q_{\wh{\Sigma}_d}\wh{\Sigma}Q_{\wh{\Sigma}_d}^{\top}-D\Big\vert\Big\}=p^{-1}\tr\Big\{\Big\vert Q_B\Sigma Q_B^{\top}-D+Q_{\wh{\Sigma}_d}\Sigma Q_{\wh{\Sigma}_d}^{\top}-Q_B\Sigma Q_B^{\top}+ Q_{\wh{\Sigma}_d}\wh{\Sigma}Q_{\wh{\Sigma}_d}^{\top}-Q_{\wh{\Sigma}_d}\Sigma Q_{\wh{\Sigma}_d}^{\top}\Big\vert\Big\}\leq W_1+W_2+W_3$, where $W_1=p^{-1}\tr\Big\{\Big\vert Q_B\Sigma Q_B^{\top}-D\Big\vert\Big\}$,  $W_2=p^{-1}\tr\Big\{\Big\vert Q_{\wh{\Sigma}_d}\Sigma Q_{\wh{\Sigma}_d}^{\top}-Q_B\Sigma Q_B^{\top}\Big\vert\Big\}$, $W_3=p^{-1}\tr\Big\{\Big\vert Q_{\wh{\Sigma}_d}\big(\wh{\Sigma}-\Sigma\big)Q_{\wh{\Sigma}_d}^{\top}\Big\vert\Big\}$, $Q_B=I_p-H_B$ and $H_B=B\big(B^{\top}B\big)^{-1}B^{\top}$. Therefore, it suffices to show that $W_1\to 0$ as $p\to\infty$, $W_2=o_p(1)$ and $W_3=o_p(1)$ as $n\to\infty$, respectively. The proof of those conclusions are given in the following three steps.

\sc{Step 1.}\normalfont$\ $We start with $W_1=p^{-1}\tr\Big\{\big\vert Q_B\Sigma Q_B^{\top}-D\big\vert\Big\}=p^{-1}\tr\Big\{\big\vert Q_BDQ_B^{\top}-D\big\vert\Big\}$ due to the fact that $\Sigma=BB^{\top}+D$ and $Q_B=I_p-H_B$. Direct computation leads to $W_1\leq W_1^a+2W_1^b$, where $W_1^a=p^{-1}\tr\Big\{\big\vert H_BDH_B^{\top}\big\vert\Big\}$ and $W_1^b=p^{-1}\tr\Big\{\big\vert DH_B\big\vert\Big\}$. Therefore, it suffices to show that both $W_1^a\to 0$ and $W_1^b\to 0$ as $p\to\infty$. Write $H_B=\big(h_1^B,\ldots,h_p^B\big)^{\top}\in\mR^{p\times p}$ with $h_j^B=\big(h_{j1}^B,\ldots,h_{jp}^B\big)^{\top}\in\mR^p$ for every $1\leq j\leq p$. Note that $H_B$ is a symmetric projection matrix such that $H_B^2=H_B$ and $h_{jj}^B\geq 0$. In the meanwhile, by condition (C3), we know that $0<\tau_{\min}\leq \tau_{j}\leq \tau_{\max}<1$ for every $1\leq j\leq p$. Then, we can directly compute $W_1^a=p^{-1}\tr\Big\{\big\vert H_BDH_B^{\top}\big\vert\Big\}=p^{-1}\sum_{j=1}^p\tau_j^2\big(h_j^B\big)^{\top}h_j^B\leq \big(\tau_{\max}^2/p\big)\tr\big(H_BH_B^{\top}\big)= \big(\tau_{\max}^2/p\big)\tr\big(H_B\big)=\tau_{\max}^2d/p \to 0$ as $p\to\infty$. For $ W_1^b$, direct computation leads to $W_1^b=p^{-1}\tr\Big\{\big\vert DH_B\big\vert\Big\}=p^{-1}\sum_{j=1}^p\vert\tau_j^2h_{jj}^B\vert\leq \big(\tau_{\max}^2/p\big)\tr\big(H_B\big)=\tau_{\max}^2d/p \to 0$ as $p\to\infty$. Therefore, we can obtain that $W_1\leq W_1^a+2W_1^b\to 0$ as $p\to\infty$.

\sc{Step 2.}\normalfont$\ $Next, we study $ W_2=p^{-1}\tr\Big\{\Big\vert Q_{\wh{\Sigma}_d}\Sigma Q_{\wh{\Sigma}_d}^{\top}-Q_B\Sigma Q_B^{\top}\Big\vert\Big\}$. We know that $ Q_{\wh{\Sigma}_d}$, $H_{\wh{\Sigma}_d}$, $\Sigma$, $Q_B$ and $H_B$ are all symmetric matrices. Recall that $Q_B=I_p-H_B$ and $Q_{\wh{\Sigma}_d}=I_p-H_{\wh{\Sigma}_d}$. We then have $\Sigma Q_B=BB^{\top}Q_B+DQ_B=DQ_B$ and $Q_B\Sigma=Q_BD$. Next, we can compute $W_2$ as $W_2=p^{-1}\tr\Big\{\Big\vert \big(Q_{\wh{\Sigma}_d}-Q_B\big)\Sigma \big(Q_{\wh{\Sigma}_d}-Q_B\big)+\big(Q_{\wh{\Sigma}_d}-Q_B\big)\Sigma Q_B+Q_B\Sigma\big(Q_{\wh{\Sigma}_d}-Q_B\big) \Big\vert\Big\}\leq W_2^a+2W_2^b+W_2^c+W_2^d$, where $W_2^a=p^{-1}\tr\Big\{\Big\vert \big(H_{\wh{\Sigma}_d}-H_B\big)\Sigma \big(H_{\wh{\Sigma}_d}-H_B\big)\Big\vert\Big\}$, $W_2^b=p^{-1}\tr\Big\{\Big\vert \big(H_{\wh{\Sigma}_d}-H_B\big)D\Big\vert\Big\}$, $W_2^c=p^{-1}\tr\Big\{\Big\vert \big(H_{\wh{\Sigma}_d}-H_B\big)DH_B\Big\vert\Big\}$ and $W_2^d=p^{-1}\tr\Big\{\Big\vert H_BD\big(H_{\wh{\Sigma}_d}-H_B\big)\Big\vert\Big\}$. To prove $W_2=o_p(1)$ as $n\to\infty$, it suffices to show that $W_2^a=o_p(1)$, $W_2^b=o_p(1)$, $W_2^c=o_p(1)$ and $W_2^d=o_p(1)$ as $n\to\infty$, respectively. The details are given below.

\sc{Step 2.1.}\normalfont$\ $We start with $W_2^a$. Recall that $\Sigma=BB^{\top}+D$ and $\Sigma_B=B^{\top}B/p=\diag\big\{\lambda_1\big(\Sigma_B\big),\ldots,\lambda_d\big(\Sigma_B\big)\big\}$. Note that $ H_{\wh{\Sigma}_d}-H_B$ is a symmetric matrix. Write $\Delta_H=H_{\wh{\Sigma}_d}-H_B=\big(\delta_1^H,\ldots,\delta_p^H\big)^{\top}\in\mR^{p\times p}$ with $\delta_j^H=\big(\delta_{j1}^H,\ldots,\delta_{jp}^H\big)^{\top}\in\mR^p$ for every $1\leq j\leq p$. Then, we can obtain that $ \mD\big(\wh{\Sigma}_d,B\big)=\tr\big(H_{\wh{\Sigma}_d}-H_B\big)^2=\tr\big(\Delta_H\Delta_H^{\top}\big)=\sum_{j_1=1}^p\sum_{j_2=1}^p\big(\delta_{j_1j_2}^H\big)^2$. By Theorem 2, we know that $ \mD\big(\wh{\Sigma}_d,B\big)=\mD\big(B,\wh{\Sigma}_d\big)=o_p(1)$ as $n\to\infty$. Immediately, we have $\tr\big(\Delta_H\Delta_H^{\top}\big)=\sum_{j_1=1}^p\sum_{j_2=1}^p\big(\delta_{j_1j_2}^H\big)^2=o_p(1)$ as $n\to\infty$. Note that $\Sigma$ is a positive definite matrix. Direct computation leads to $ W_2^a=p^{-1}\tr\Big\{\big\vert\Delta_H^{\top}\Sigma\Delta_H\big\vert\Big\}=p^{-1}\sum_{j=1}^p\big(\delta_j^H\big)^{\top}\Sigma\delta_j^H\leq p^{-1}\lambda_{\max}\big(\Sigma\big)\sum_{j=1}^p\big(\delta_j^H\big)^{\top}\delta_j^H$. In the meanwhile, we can obtain that $ p^{-1}\lambda_{\max}\big(\Sigma\big)\leq \lambda_{\max}\big(BB^{\top}/p\big)+\lambda_{\max}\big(D\big)/p\leq \lambda_1\big(\Sigma_B\big)+\tau_{\max}^2/p$. We then have $W_2^a\leq \big\{\lambda_1\big(\Sigma_B\big)+\tau_{\max}^2/p\big\}\tr\big(\Delta_H\Delta_H^{\top}\big) =o_p(1)$ as $n\to\infty$. This completes the proof of $W_2^a=o_p(1)$ as $n\to\infty$.

\sc{Step 2.2.}\normalfont$\ $We next study $ W_2^b$, $W_2^c$ and $W_2^d$. Recall that $ W_2^b=p^{-1}\tr\Big\{\Big\vert \big(H_{\wh{\Sigma}_d}-H_B\big)D\Big\vert\Big\}$, $ D=\diag\big\{\tau_{1}^2,\ldots,\tau_{p}^2\big\}$ and $\tau_{j}^2\leq \tau_{\max}^2<1$ for every $1\leq j\leq p$. Direct computation leads to $ W_2^b=p^{-1}\sum_{j=1}^{p}\big\vert\delta_{jj}^H\tau_{j}^2\big\vert\leq \tau_{\max}^2p^{-1}\sum_{j=1}^{p}\big\vert\delta_{jj}^H\big\vert\leq p^{-1}\sum_{j_1=1}^p\sum_{j_2=1}^p\big\vert\delta_{j_1j_2}^H\big\vert$. By Cauchy-Schwarz inequality \citep{hardy1934}, we can obtain that $ \sum_{j_1=1}^p\sum_{j_2=1}^p$ $\big\vert\delta_{j_1j_2}^H\big\vert\leq \Big\{p^2\sum_{j_1=1}^p\sum_{j_2=1}^p$ $\big(\delta_{j_1j_2}^H\big)^2\Big\}^{1/2} $. By \sc{Step 2.1,}\normalfont$\ $we know that $ \sum_{j_1=1}^p\sum_{j_2=1}^p\big(\delta_{j_1j_2}^H\big)^2$ $=o_p(1)$ as $n\to\infty$. Therefore, we have $W_2^b\leq p^{-1}\sum_{j_1=1}^p$ $\sum_{j_2=1}^p\big\vert\delta_{j_1j_2}^H\big\vert\leq \Big\{\sum_{j_1=1}^p\sum_{j_2=1}^p\big(\delta_{j_1j_2}^H\big)^2\Big\}^{1/2}=o_p(1)$ as $n\to\infty$. Following the previous notation, we can compute $W_2^c$ as $W_2^c=p^{-1}\tr\Big\{\Big\vert \big(H_{\wh{\Sigma}_d}-H_B\big)DH_B\Big\vert\Big\}=p^{-1}\sum_{j=1}^p\Big\vert\tau_j^2\big(\delta_j^H\big)^{\top}h_j^B\Big\vert\leq \tau_{\max}^2\mT^H$, where $\mT^H=p^{-1}$ $\tr\Big\{\Big\vert\big(H_{\wh{\Sigma}_d}-H_B\big)H_B\Big\vert\Big\}$. Similarly, we can also calculate $W_2^d=p^{-1}\tr\Big\{\Big\vert H_BD\big(H_{\wh{\Sigma}_d}-H_B\big)\Big\vert\Big\}=p^{-1}\sum_{j=1}^p\big\vert\tau_j^2\big(h_j^B\big)^{\top}\delta_j^H\big\vert\leq \tau_{\max}^2p^{-1}\tr\Big\{\Big\vert H_B\big(H_{\wh{\Sigma}_d}-H_B\big)\Big\vert\Big\} =\tau_{\max}^2\mT^H$. Therefore, it suffices to show that $\mT^H=o_p(1)$ as $n\to\infty$. Since $H_B$ is a projection matrix, we can obtain that $\big\vert h_{j_1j_2}^B\big\vert\leq 1$ for every $1\leq j_1,j_2\leq p$ \citep{mohammadi2016}. Then, direct computation leads to $\mT^H=p^{-1}\sum_{j_1=1}^p\Big\vert\sum_{j_2=1}^p\delta_{j_1j_2}^H$ $h_{j_1j_2}^B\Big\vert\leq p^{-1}\sum_{j_1=1}^p\sum_{j_2=1}^p\big\vert\delta_{j_1j_2}^H\big\vert$. By the proof of $W_2^b$, we have $ p^{-1}\sum_{j_1=1}^p\sum_{j_2=1}^p\big\vert\delta_{j_1j_2}^H\big\vert=o_p(1)$ as $n\to\infty$. Thus, we can directly obtain that $\mT^H=o_p(1)$ as $n\to\infty$. This further implies that $W_2^c=o_p(1)$ and $W_2^d=o_p(1)$ as $n\to\infty$. Combining the results of \sc{Step 2.1,}\normalfont$\ $we have $W_2\leq W_2^a+2W_2^b+W_2^c+W_2^d=o_p(1)$ as $n\to\infty$. This completes the proof of \sc{Step 2.}\normalfont$\ $

\sc{Step 3.}\normalfont$\ $Lastly, we prove $ W_3=o_p(1)$ as $n\to\infty$ and the theorem conclusion. Recall that $W_3=p^{-1}\tr\Big\{\Big\vert Q_{\wh{\Sigma}_d}\big(\wh{\Sigma}-\Sigma\big)Q_{\wh{\Sigma}_d}^{\top}\Big\vert\Big\}$. Note that $Q_{\wh{\Sigma}_d}=I_p-H_{\wh{\Sigma}_d}$ is a symmetric projection  matrix such that $Q_{\wh{\Sigma}_d}^2=Q_{\wh{\Sigma}_d}$ and $\tr\big(Q_{\wh{\Sigma}_d}\big)=p-d$. Write $Q_{\wh{\Sigma}_d}=\big(q_1,\ldots,q_p\big)^{\top}\in\mR^{p\times p}$. We can directly compute $ W_3$ as $W_3=p^{-1}\sum_{j=1}^p\big\vert q_j^{\top}\big(\wh{\Sigma}-\Sigma\big)q_j\big\vert\leq \big\Vert\wh{\Sigma}-\Sigma\big\Vert\sum_{j=1}^pq_j^{\top}q_j/p=\big\Vert\wh{\Sigma}-\Sigma\big\Vert\tr\big(Q_{\wh{\Sigma}_d}Q_{\wh{\Sigma}_d}^{\top}\big)/p=\big\Vert\wh{\Sigma}-\Sigma\big\Vert\big(p-d\big)/p\leq \big\Vert\wh{\Sigma}-\Sigma\big\Vert$. By \sc{Step 4}\normalfont$\ $in the proof of Theorem 2, we know that $ \big\Vert\wh{\Sigma}-\Sigma\big\Vert\xrightarrow{p}0$ as $n\to\infty$. Therefore, we have $ W_3=o_p(1)$ as $n\to\infty$. This finishes the proof of \sc{Step 3}\normalfont$\ $and also completes the entire theorem proof.

\subsection*{\textbf{Appendix A.4. Proof of Theorem 4}}

Consider one particular specification about $B$ and $Z_i$ for model (2.4). The theorem conclusion can be proved in a total of four  steps. In the first step, we show that $\big\Vert\wt{Z}_i-Z_i\big\Vert=O_p(1/\sqrt{p})$, where $\wt{Z}_i=\argmax_{Z_i}\wt{\mL}_{B}^{(i)}\big(Z_i\big) $ and
\begin{align}\label{b4.1}
	\wt{\mL}_{B}^{(i)}\big(Z_i\big)=&\frac{1}{p}\sum_{j=1}^p\Bigg(Y_{ij}\log\Phi\Big\{\Big(b_j^{\top}Z_i-c_j\Big)/\tau_j\Big\}\nonumber\\
	&+\Big(1-Y_{ij}\Big)\log\bigg[1-\Phi\Big\{\Big(b_j^{\top}Z_i-c_j\Big)/\tau_j\Big\}\bigg]\Bigg)
\end{align} 
is the oracle log-likelihood function in the sense that both $c_i$ and $\tau_j$ are given. In the second step, we prove that $ \sup_{\Vert z\Vert\leq Z_{\max}}\big\vert\mL_{B,\tau}^{(i)}\big(z\big)-\wt{\mL}_{B}^{(i)}\big(z\big)\big\vert=o_p(1)$, where $\mL_{B,\tau}^{(i)}\big(z\big)=p^{-1}\sum_{j=1}^pI\big(\wh{\tau}_j>\tau\big)\Big(Y_{ij}\log\Phi\big\{\big(b_j^{\top}z-\wh{c}_j\big)/\wh{\tau}_j\big\}+\big(1-Y_{ij}\big)\log\Big[1$ $-\Phi\big\{\big(b_j^{\top}z-\wh{c}_j\big)/\wh{\tau}_j\big\}\Big]\Big)$. In the third step, we prove that $\Big\vert\mL_{B,\tau}^{(i)}\big(\wh{Z}^*_i\big)-\mL_{B,\tau}^{(i)}\big(\wt{Z}_i^*\big)\Big\vert=o_p(1)$, where $\wh{Z}^*_i=\argmax_{Z_i}$ $\mL_{B,\tau}^{(i)}\big(Z_i\big) $, $\wt{Z}^*_i=\wh{\Omega}\wh{Z}_i$, $\wh{Z}_i=\argmax_{Z_i}\mL_{\wh{B},\tau}^{(i)}\big(Z_i\big) $, and $\wh{\Omega}=\big(B^{\top}B\big)^{-1}B^{\top}\wh{B}\in\mR^{d\times d}$. In the last step, we show that $ p^{-1/2}\big\Vert\wh{B}\wh{Z}_i-BZ_i\big\Vert \xrightarrow{p}0$ as $n\to\infty$. The details are given below.

\sc{Step 1.}\normalfont$\ $We start with $\big\Vert\wt{Z}_i-Z_i\big\Vert$. Following \cite{Fan2001}, we would like to show that for any given $\epsilon>0$, there exists a sufficiently large constant $ C_{\epsilon} $ such that
\begin{equation}\label{b4.2}
	\mathop{\text{lim inf}}\limits_{n\to\infty} P\bigg[\sup\limits_{\lVert u\lVert=1} p\Big\{\wt{\mL}^{(i)}_B\Big(Z_i+uC_{\epsilon}/\sqrt{p}\Big)-\wt{\mL}^{(i)}_B\big(Z_i\big)\Big\}<0\bigg]\geq 1-\epsilon.
\end{equation}
To this end, we apply the Taylor's expansion about $ \Delta^{(i)}= p\wt{\mL}^{(i)}_B\big(Z_i+uC_{\epsilon}/\sqrt{p}\big)-p\wt{\mL}^{(i)}_B\big(Z_i\big)$ at $ Z_i $ as $ \Delta^{(i)}=\Big\{p^{1/2}\dot{\wt{\mL}}^{(i)}_B\big(Z_i\big)\Big\}^{\top}uC_{\epsilon} -2^{-1}C_{\epsilon}^2u^{\top}\Big\{-\ddot{\wt{\mL}}^{(i)}_B\big(Z_i\big)\Big\}u+o_p(1)$, where $ \dot{\wt{\mL}}^{(i)}_B\big(Z_i\big)=\partial\wt{\mL}^{(i)}_B\big(Z_i\big)/\partial Z_i\in\mR^d $ and $ \ddot{\wt{\mL}}^{(i)}_B\big(Z_i\big)=\partial^2\wt{\mL}^{(i)}_B\big(Z_i\big)/\partial Z_i\partial Z_i^{\top}\in\mR^{d\times d}$ are the 1st and 2nd order partial derivatives of $ \wt{\mL}^{(i)}_B\big(Z_i\big) $ with respect to $ Z_i$, respectively.  Then, we have $\sup_{\Vert u\Vert=1}\Delta^{(i)}\leq \mQ_{11}-\mQ_{12}+o_p(1)$, where $\mQ_{11}=C_{\epsilon}p^{1/2}\Big\Vert \dot{\wt{\mL}}^{(i)}_B\big(Z_i\big)\Big\Vert$ and $\mQ_{12}=2^{-1}C_{\epsilon}^2\lambda_{\min}\Big\{-\ddot{\wt{\mL}}^{(i)}_B\big(Z_i\big)\Big\}$. Following the proof of Theorem 1 in \cite{Fan2001}, we can obtain that $ p^{1/2}\Big\Vert\dot{\wt{\mL}}^{(i)}_B\big(Z_i\big)\Big\Vert=O_p(1)$. The verification details are given in \sc{Step 1.1}\normalfont$\ $in Appendix B.2. Immediately, we know that $\mQ_{11}=O_p(1)$. Let $ I(Z_i)=-E\Big\{\ddot{\wt{\mL}}^{(i)}_B\big(Z_i\big)\Big\}$ be the Fisher information matrix of the oracle log-likelihood function $\wt{\mL}^{(i)}_B\big(Z_i\big)$. Direct computation leads to
\begin{equation}\label{b4.3}
	I(Z_i)=\dfrac{1}{p}\sum_{j=1}^p\dfrac{b_jb_j^{\top}\phi^2(x_{ij})}{\tau_j^2\Phi(x_{ij})\big\{1-\Phi(x_{ij})\big\}}
\end{equation}
where $ \phi(x)=(2\pi)^{-1/2}$  $\exp(-x^2/2)$ is the probability density function of a standard normal distribution and $x_{ij}=\big(b_j^{\top}Z_i-c_j\big)/\tau_j$. By \sc{Step 2}\normalfont$\ $of Theorem 1 and condition (C5), we have $ \tau_{\min}^2\leq \tau_{j}^2\leq \tau_{\max}^2$ and $\vert x_{ij}\vert\leq x_{\max}$, where $x_{\max}=\tau_{\min}^{-1}\big(Z_{\max}+c_{\max}\big)$. Consequently, $I(Z_i)$ is a positive definite matrix with $\lambda_{\min}\big\{I(Z_i)\big\}\geq \nu_{\min}>0$ for some constant $\nu_{\min}$. Write $-\ddot{\wt{\mL}}^{(i)}_B\big(Z_i\big)=I(Z_i)+\wt{\Delta}^{(i)}$ with $ \wt{\Delta}^{(i)}=-\ddot{\wt{\mL}}^{(i)}_B\big(Z_i\big)-I(Z_i)$. By Law of Numbers, we know that $ \big\Vert\wt{\Delta}^{(i)}\big\Vert=o_p(1)$. Therefore, we have $\mQ_{12}\geq 2^{-1}C_{\epsilon}^2\nu_{\min}\big\{1+o_p(1)\big\}$. By choosing
a sufficiently large $C_{\epsilon}$, $\mQ_{12}$ can dominate $\mQ_{11}$. As a result, (\ref{b4.2}) holds for  a sufficiently large $C_{\epsilon}$. This completes the proof of $\big\Vert\wt{Z}_i-Z_i\big\Vert=O_p(1/\sqrt{p})$.

\sc{Step 2.}\normalfont$\ $Next, we study  $\sup_{\Vert z\Vert\leq Z_{\max}}\big\vert\mL_{B,\tau}^{(i)}(z)-\wt{\mL}_{B}^{(i)}(z)\big\vert$. Note that $I(\tau_j>\tau)=1$ for every $1\leq j\leq p$ since $0<\tau<\tau_{\min}$. We then have $\wt{\mL}_{B}^{(i)}(z)=\wt{\mL}_{B,\tau}^{(i)}(z)$, where $ \wt{\mL}_{B,\tau}^{(i)}(z)=p^{-1}\sum_{j=1}^pI\big(\tau_j>\tau\big)\Big(Y_{ij}\log\Phi\big\{\big(b_j^{\top}z-c_j\big)/\tau_j\big\}+\big(1-Y_{ij}\big)\log\Big[1-\Phi\big\{\big(b_j^{\top}z-c_j\big)/\tau_j\big\}\Big]\Big)$. Thus, $ \mL_{B,\tau}^{(i)}(z)$ is almost the same as $ \wt{\mL}_{B,\tau}^{(i)}(z)$ but with $\big(c_j,\tau_j\big)$ replaced by $\big(\wh{c}_j,\wh{\tau}_j\big)$. Define $\zeta_j=\big(c_j,\tau_j^2\big)^{\top}\in\mR^2$. Write $ \mL_{B,\tau}^{(i)}(z)=p^{-1}\sum_{j=1}^pI\big(\wh{\tau}_j>\tau\big)\ell_{B,\tau}^{(i,j)}\big(z,\wh{\zeta}_j\big)$ and $ \wt{\mL}_{B,\tau}^{(i)}(z)=p^{-1}\sum_{j=1}^pI\big(\tau_j>\tau\big)\ell_{B,\tau}^{(i,j)}\big(z,\zeta_j\big)$ with $ \ell_{B,\tau}^{(i,j)}\big(z,\zeta_j\big)=\Big(Y_{ij}\log\Phi\big\{\big(b_j^{\top}z-c_j\big)/\tau_j\big\}+\big(1-Y_{ij}\big)\log\Big[1-\Phi\big\{\big(b_j^{\top}z-c_j\big)/\tau_j\big\}\Big]\Big)$. Then, we conduct the Taylor's expansion about $\Delta^{(i)}_{\mL}=\mL_{B,\tau}^{(i)}(z)-\wt{\mL}_{B,\tau}^{(i)}(z)$ for $\wh{\zeta}_j $ at $\zeta_j $ as $ \Delta^{(i)}_{\mL}=\mQ_{21}+\mQ_{22}$, where $\mQ_{21}=-p^{-1}\sum_{j=1}^pI\big(\wh{\tau}_j\leq\tau\big)\ell_{B,\tau}^{(i,j)}\big(z,\zeta_j\big)$ and $\mQ_{22}=p^{-1}\sum_{j=1}^pI\big(\wh{\tau}_j>\tau\big)\big\{\dot{\ell}_{B,\tau}^{(i,j)}\big(z,\wt{\zeta}_j\big)\big\}^{\top}\big(\wh{\zeta}_j-\zeta_j\big)$. Here $ \dot{\ell}_{B,\tau}^{(i,j)}\big(z,\zeta_j\big)=\partial\ell_{B,\tau}^{(i,j)}\big(z,\zeta_j\big)/\partial \zeta_j\in\mR^2$ is the first-order partial derivative of $ \ell_{B,\tau}^{(i,j)}\big(z,\zeta_j\big)$ with respect to $\zeta_j$ and $ \wt{\zeta}_j=\alpha_j\wh{\zeta}_j+(1-\alpha_j)\zeta_j$ for some $0<\alpha_j<1$. We then have $\sup_{\Vert z\Vert\leq Z_{\max}}\big\vert\Delta^{(i)}_{\mL}\big\vert\leq \sup_{\Vert z\Vert\leq Z_{\max}}\max_j\vert \mQ_{21}\vert+\sup_{\Vert z\Vert\leq Z_{\max}}\max_j\vert\mQ_{22}\vert$. One can verify that $ \sup_{\Vert z\Vert\leq Z_{\max}}\max_j\vert\mQ_{21}\vert=o_p(1)$ as $n\to\infty$ and $\sup_{\Vert z\Vert\leq Z_{\max}}\max_j\vert\mQ_{22}=o_p(1)$ as $n\to\infty$. The verification details are shown in \sc{Step 2.1-2.2}\normalfont$\ $ in Appendix B.2. Therefore, we can obtain that $ \sup_{\Vert z\Vert\leq Z_{\max}}\big\vert\Delta^{(i)}_{\mL}\big\vert=o_p(1)$. 

\sc{Step 3.}\normalfont$\ $We next prove that $\big\vert\mL_{B,\tau}^{(i)}\big(\wh{Z}_i^*\big)-\mL_{B,\tau}^{(i)}\big(\wt{Z}_i^*\big)\big\vert=o_p(1)$. Recall that $\wh{Z}_i^*=\argmax_{Z_i}\mL_{B,\tau}^{(i)}\big(Z_i\big) $, $\wt{Z}_i^*=\wh{\Omega}\wh{Z}_i$, $ \wh{Z}_i=\argmax_{Z_i}\mL_{\wh{B},\tau}^{(i)}\big(Z_i\big) $ and $\wh{\Omega}=\big(B^{\top}B\big)^{-1}B^{\top}\wh{B}$. Note that the key difference between $ \wh{Z}_i $ and $ \wh{Z}_i^* $ is due to the relationship between $ \wh{B}$ and $ B $. Recall that $\wh{B}=\wh{\Sigma}_d\wh{\Lambda}_d^{1/2}$ and $H_{\wh{B}}=\wh{B}\big(\wh{B}^{\top}\wh{B}\big)^{-1}\wh{B}^{\top}$. We then have $H_{\wh{B}}=\wh{\Sigma}_d\wh{\Lambda}_d^{1/2}\big(\wh{\Lambda}_d^{1/2}\wh{\Sigma}_d^{\top}\wh{\Sigma}_d\wh{\Lambda}_d^{1/2}\big)^{-1}\wh{\Lambda}_d^{1/2}\wh{\Sigma}_d=\wh{\Sigma}_d\wh{\Sigma}_d^{\top}=H_{\wh{\Sigma}_d}$. Write $ \wh{B}=H_{\wh{B}}\wh{B}=H_B\wh{B}+\big(H_{\wh{B}}-H_B\big)\wh{B}=B\wh{\Omega}+\Delta_H\wh{B}$, where $ \Delta_H=H_{\wh{\Sigma}_d}-H_B=\big(\delta_1^H,\ldots,\delta_p^H\big)^{\top} \in\mR^{p\times p}$ and $\delta_j^H=\big(\delta_{j1}^H,\ldots,\delta_{jp}^H\big)^{\top}\in\mR^p$ for every $1\leq j\leq p$. Recall that $ B=\big(b_1,\ldots,b_p\big)^{\top}\in\mR^{p\times d}$ and $ \wh{B}=\big(\wh{b}_1,\ldots,\wh{b}_p\big)^{\top}\in\mR^{p\times d}$. Then, we have $ \wh{b}_j^{\top}=b_j^{\top}\wh{\Omega}+\wh{\omega}_j^{\top}$, where $\wh{\omega}_j^{\top}=\big(\delta_j^H\big)^{\top}\wh{B}\in\mR^{1\times d} $. The desired conclusion can be established in a total of three steps. In \sc{Step 3.1,}\normalfont$\ $we show that $ p^{-1}\sum_{j=1}^p\big\Vert\wh{\omega}_j\big\Vert =o_p(1)$ as $n\to\infty$. In \sc{Step 3.2,}\normalfont$\ $we prove that $ \mL_{\wh{B},\tau}^{(i)}\big(\wh{Z}_i\big)= \mL_{B,\tau}^{(i)}\big(\wt{Z}_i^*\big)+o_p(1)$. In \sc{Step 3.3,}\normalfont$\ $we prove that $\mL_{\wh{B},\tau}^{(i)}\big(\wh{Z}_i\big)\geq \mL_{B,\tau}^{(i)}\big(\wh{Z}_i^*\big)+o_p(1)$. Therefore, we have $ \mL_{B,\tau}^{(i)}\big(\wt{Z}_i^*\big)\geq \mL_{B,\tau}^{(i)}\big(\wh{Z}_i^*\big)+o_p(1) $. In the meanwhile, by definition, we can obtain that $ \mL_{B,\tau}^{(i)}\big(\wh{Z}_i^*\big)\geq \mL_{B,\tau}^{(i)}\big(\wt{Z}_i^*\big)$. This leads to the conclusion $\big\vert\mL_{B,\tau}^{(i)}\big(\wt{Z}_i^*\big)-\mL_{B,\tau}^{(i)}\big(\wh{Z}_i^*\big)\big\vert=o_p(1)$. The details are given below.

\sc{Step 3.1.}\normalfont$\ $We start with $  p^{-1}\sum_{j=1}^p\big\Vert\wh{\omega}_j\big\Vert$. Direct computation leads to $ p^{-1}\sum_{j=1}^p\big\Vert\wh{\omega}_j\big\Vert^2=p^{-1}\sum_{j=1}^p\wh{\omega}_j^{\top}\wh{\omega}_j=p^{-1}\tr\big(\Delta_H$ $\wh{B}\wh{B}^{\top}\Delta_H^{\top}\big)=p^{-1}\tr\big(\wh{B}\wh{B}^{\top}\Delta_H^{\top}\Delta_H\big)$. Note that both $ \wh{B}\wh{B}^{\top} $ and $ \Delta_H\Delta_H^{\top} $ are semi-positive definite matrices. By Lemma 2 in \cite{wang2012}, we can obtain that $ p^{-1}\tr\big(\wh{B}\wh{B}^{\top}\Delta_H\Delta_H^{\top}\big)\leq \lambda_{\max}\big(\wh{B}\wh{B}^{\top}/p\big)\tr\big(H_{\wh{\Sigma}_d}-H_B\big)^2=\lambda_{\max}\big(\wh{B}\wh{B}^{\top}/p\big)\mD\big(B,\wh{\Sigma}_d\big)$. By Theorem 2, we know that $\mD\big(B,\wh{\Sigma}_d\big)=o_p(1) $ as $ n\to\infty $. Recall that $\wh{B}=\wh{\Sigma}_d\wh{\Lambda}_d^{1/2}$. Then, we have $ \lambda_{\max}\big(\wh{B}\wh{B}^{\top}/p\big)=\lambda_{\max}\big(\wh{\Sigma}_d\wh{\Lambda}_d\wh{\Sigma}_d^{\top}/p\big)\leq\lambda_{\max}\big(\wh{\Sigma}/p\big)\leq p^{-1}\lambda_1\big(\Sigma\big) +p^{-1}\big\Vert\wh{\Sigma}-\Sigma\big\Vert$. By \sc{Step 4}\normalfont$\ $of Appendix B.2, we know that $ \big\Vert\wh{\Sigma}-\Sigma\big\Vert=o_p(1)$ as $n\to\infty$. Consequently, we can obtain that $ \lambda_{\max}\big(\wh{B}\wh{B}^{\top}/p\big)\leq \lambda_1\big(\Sigma_B\big)+\tau_{\max}^2+o_p(1)$ and then $p^{-1}\sum_{j=1}^p\big\Vert\wh{\omega}_j\big\Vert^2 \xrightarrow{p}0$ as $n\to\infty$. By Cauchy-Schwarz inequality \citep{hardy1934}, we have $  p^{-1}\sum_{j=1}^p\big\Vert\wh{\omega}_j\big\Vert\leq \sqrt{\sum_{j=1}^p\big\Vert\wh{\omega}_j\big\Vert^2\sum_{j=1}^pp^{-2}}=\sqrt{p^{-1}\sum_{j=1}^p\big\Vert\wh{\omega}_j\big\Vert^2}=o_p(1)$ as $n\to\infty$. This completes the proof of $  p^{-1}\sum_{j=1}^p\big\Vert\wh{\omega}_j\big\Vert=o_p(1)$ as $n\to\infty$.

\sc{Step 3.2.}\normalfont$\ $Next, we show that $ \mL_{\wh{B},\tau}^{(i)}\big(\wh{Z}_i\big)= \mL_{B,\tau}^{(i)}\big(\wt{Z}_i^*\big)+o_p(1)$. Recall that $\wt{Z}_i^*=\wh{\Omega}\wh{Z}_i$ and $ \wh{b}_j^{\top}=b_j^{\top}\wh{\Omega}+\wh{\omega}_j^{\top}$. Immediately, we can obtain that $ \wh{b}_j^{\top}\wh{Z}_i=b_j^{\top}\wt{Z}_i^*+\wh{\omega}_j^{\top}\wh{Z}_i$. By definition, we have
\begin{align}\label{b4.4}
	\mL_{\wh{B},\tau}^{(i)}\big(\wh{Z}_i\big)=&\frac{1}{p}\sum_{j=1}^pI\Big(\wh{\tau}_j>\tau\Big)\Bigg(Y_{ij}\log\Phi\Big\{\Big(b_j^{\top}\wt{Z}_i^*+\wh{\omega}_j^{\top}\wh{Z}_i-\wh{c}_j\Big)/\wh{\tau}_j\Big\}\nonumber\\
	&+\Big(1-Y_{ij}\Big)\log\bigg[1-\Phi\Big\{\Big(b_j^{\top}\wt{Z}_i^*+\wh{\omega}_j^{\top}\wh{Z}_i-\wh{c}_j\Big)/\wh{\tau}_j\Big\}\bigg]\Bigg).
\end{align}
Note that $ \dot{\Phi}(x)=\partial\Phi(x)/\partial x=\phi(x)$. Then, we can conduct the Taylor's expansion about $ \mL_{\wh{B},\tau}^{(i)}\big(\wh{Z}_i\big)$ at $ b_j^{\top}\wt{Z}_i^*$ as $ \mL_{\wh{B},\tau}^{(i)}\big(\wh{Z}_i\big)=\mQ_{32}^a+\mQ_{32}^b$, where $\mQ_{32}^a=p^{-1}\sum_{j=1}^pI\big(\wh{\tau}_j>\tau\big)\Big(Y_{ij}\log\Phi\big\{\big(b_j^{\top}\wt{Z}_i^*-\wh{c}_j\big)/\wh{\tau}_j\big\}+\big(1-Y_{ij}\big)\log\Big[1-\Phi\big\{\big(b_j^{\top}\wt{Z}_i^*-\wh{c}_j\big)/\wh{\tau}_j\big\}\Big]\Big)=\mL_{B,\tau}^{(i)}\big(\wt{Z}_i^*\big)$ and $ \mQ_{32}^b=p^{-1}\sum_{j=1}^pI\big(\wh{\tau}_j>\tau\big)\big(\wh{\omega}_j^{\top}\wh{Z}_i\big)\wh{\tau}_j^{-1}\Big[Y_{ij}-\Phi\Big\{\big(b_j^{\top}\wt{Z}_i^*+\alpha_{ij}\wh{\omega}_j^{\top}\wh{Z}_i-\wh{c}_j\big)/\wh{\tau}_j\Big\}\Big]\phi\Big\{\big(b_j^{\top}\wt{Z}_i^*+\alpha_{ij}\wh{\omega}_j^{\top}\wh{Z}_i-\wh{c}_j\big)/\wh{\tau}_j\Big\}\big/\Big(\Phi\Big\{\big(b_j^{\top}\wt{Z}_i^*+\alpha_{ij}\wh{\omega}_j^{\top}\wh{Z}_i-\wh{c}_j\big)/\wh{\tau}_j\Big\}\Big[1-\Phi\Big\{\big(b_j^{\top}\wt{Z}_i^*+\alpha_{ij}\wh{\omega}_j^{\top}\wh{Z}_i-\wh{c}_j\big)/\wh{\tau}_j\Big\}\Big]\Big)$ for some $ 0<\alpha_{ij}<1 $. Therefore, we have $\mL_{\wh{B},\tau}^{(i)}\big(\wh{Z}_i\big)= \mL_{B,\tau}^{(i)}\big(\wt{Z}_i^*\big)+\mQ_{32}^b$. Next, it suffices to prove that $ \mQ_{32}^b=o_p(1) $. Direct computation leads to $ \mQ_{32}^b=p^{-1}\sum_{j=1}^pI\big(\wh{\tau}_j>\tau\big)\big(\wh{\omega}_j^{\top}\wh{Z}_i\big)\wh{\tau}_j^{-1}\big\{Y_{ij}-\Phi(x_{ij}^{\alpha})\big\}\phi(x_{ij}^{\alpha})/\big[\Phi(x_{ij}^{\alpha})\big\{1- \Phi(x_{ij}^{\alpha})\big\}\big]$, where $x_{ij}^{\alpha}=\big(b_j^{\top}\wt{Z}_i^*+\alpha_{ij}\wh{\omega}_j^{\top}\wh{Z}_i-\wh{c}_j\big)/\wh{\tau}_j$. Note that $ Y_{ij}\in\big\{0,1\big\} $ and $ 0\leq \Phi(x)\leq 1 $ for every $x\in\mR$. Then, we have $ \big\vert\mQ_{32}^b\big\vert\leq 2\big\Vert\wh{Z}_i\big\Vert\max_j\Big\vert I\big(\wh{\tau}_j>\tau\big)\wh{\tau}_j^{-1}\phi(x_{ij}^{\alpha})/\big[\Phi(x_{ij}^{\alpha})\big\{1- \Phi(x_{ij}^{\alpha})\big\}\big]\Big\vert\cdot p^{-1}\sum_{j=1}^p\big\Vert\wh{\omega}_j\big\Vert$. By \sc{Step 3.1,}\normalfont$\ $we know that $ p^{-1}\sum_{j=1}^p\big\Vert\wh{\omega}_j\big\Vert\to 0$ as $n\to\infty$. It can be verified that $\big\Vert\wh{Z}_i\big\Vert\max_j\Big\vert I\big(\wh{\tau}_j>\tau\big)\wh{\tau}_j^{-1}\phi(x_{ij}^{\alpha})/\big[\Phi(x_{ij}^{\alpha})\big\{1- \Phi(x_{ij}^{\alpha})\big\}\big]\Big\vert= O_p(1)$. The verification details of those conclusions are given in \sc{Step 3.2.1}\normalfont$\ $in Appendix B.2. Consequently, $\mQ_{32}^b=o_p(1)$ is proved. This completes the proof of $ \mL_{\wh{B},\tau}^{(i)}\big(\wh{Z}_i\big)=\mL_{B,\tau}^{(i)}\big(\wt{Z}_i^*\big)+o_p(1)$.

\sc{Step 3.3.}\normalfont$\ $We next prove that $ \mL_{\wh{B},\tau}^{(i)}\big(\wh{Z}_i\big)\geq \mL_{B,\tau}^{(i)}\big(\wh{Z}_i^*\big)+o_p(1) $. By definition, we know that $\mL_{\wh{B},\tau}^{(i)}\big(\wh{Z}_i\big)\geq \mL_{\wh{B},\tau}^{(i)}\big(\wh{\Omega}^{-1}\wh{Z}_i^*\big)$. Recall that $ \wh{b}_j^{\top}=b_j^{\top}\wh{\Omega}+\wh{\omega}_j^{\top}$. Direct computation leads to 
\begin{align}\label{b4.5}
	\mL_{\wh{B},\tau}^{(i)}\Big(\wh{\Omega}^{-1}\wh{Z}_i^*\Big)=&\frac{1}{p}\sum_{j=1}^pI\Big(\wh{\tau}_j>\tau\Big)\Bigg(Y_{ij}\log\Phi\Big\{\Big(b_j^{\top}\wh{Z}_i^*+\wh{\omega}_j^{\top}\wh{\Omega}^{-1}\wh{Z}_i^*-\wh{c}_j\Big)/\wh{\tau}_j\Big\}\nonumber\\
	&+\Big(1-Y_{ij}\Big)\log\bigg[1-\Phi\Big\{\Big(b_j^{\top}\wh{Z}_i^*+\wh{\omega}_j^{\top}\wh{\Omega}^{-1}\wh{Z}_i^*-\wh{c}_j\Big)/\wh{\tau}_j\Big\}\bigg]\Bigg).
\end{align}
Similar as before, we conduct the Taylor's expansion about $ \mL_{\wh{B},\tau}^{(i)}\Big(\wh{\Omega}^{-1}\wh{Z}_i^*\Big) $ at $ b_j^{\top}\wh{Z}_i^*$ as $ \mL_{\wh{B},\tau}^{(i)}\big(\wh{\Omega}^{-1}\wh{Z}_i^*\big)=\mQ_{33}^a+\mQ_{33}^b$, where $\mQ_{33}^a=p^{-1}\sum_{j=1}^pI\big(\wh{\tau}_j>\tau\big)\Big(Y_{ij}\log\Phi\big\{\big(b_j^{\top}\wh{Z}_i^*-\wh{c}_j\big)/\wh{\tau}_j\big\}+\big(1-Y_{ij}\big)\log\Big[1-\Phi\big\{\big(b_j^{\top}\wh{Z}_i^*-\wh{c}_j\big)/\wh{\tau}_j\big\}\Big]\Big)=\mL_{B,\tau}^{(i)}\big(\wh{Z}_i^*\big)$ and $ \mQ_{33}^b=p^{-1}\sum_{j=1}^pI\big(\wh{\tau}_j>\tau\big)\Big(\wh{\omega}_j^{\top}$ $\wh{\Omega}^{-1}\wh{Z}_i^*\Big)\wh{\tau}_j^{-1}\Big[Y_{ij}-\Phi\Big\{\big(b_j^{\top}\wh{Z}_i^*+\eta_{ij}\wh{\omega}_j^{\top}\wh{\Omega}^{-1}\wh{Z}_i^*-\wh{c}_j\big)/\wh{\tau}_j\Big\}\Big]\phi\Big\{\big(b_j^{\top}\wh{Z}_i^*+\eta_{ij}\wh{\omega}_j^{\top}\wh{\Omega}^{-1}\wh{Z}_i^*-\wh{c}_j\big)/\wh{\tau}_j\Big\}\big/$ $\Big(\Phi\Big\{\big(b_j^{\top}\wh{Z}_i^*+\eta_{ij}\wh{\omega}_j^{\top}\wh{\Omega}^{-1}\wh{Z}_i^*-\wh{c}_j\big)/\wh{\tau}_j\Big\}\Big[1-\Phi\Big\{\big(b_j^{\top}\wh{Z}_i^*+\eta_{ij}\wh{\omega}_j^{\top}\wh{\Omega}^{-1}\wh{Z}_i^*-\wh{c}_j\big)/\wh{\tau}_j\Big\}\Big]\Big) $ for some $ 0<\eta_{ij}<1 $. Therefore, we have $ \mL_{\wh{B},\tau}^{(i)}\big(\wh{Z}_i\big)\geq \mL_{B,\tau}^{(i)}\big(\wh{Z}_i^*\big)+\mQ_{33}^b$. Next, it suffices to prove that $ \mQ_{33}^b=o_p(1) $. Direct computation leads to $ \mQ_{33}^b=p^{-1}\sum_{j=1}^pI\big(\wh{\tau}_j>\tau\big)\Big(\wh{\omega}_j^{\top}\wh{\Omega}^{-1}\wh{Z}_i^*\Big)\wh{\tau}_j^{-1}\big\{Y_{ij}-\Phi\big(x_{ij}^{\eta}\big)\big\}\phi\big(x_{ij}^{\eta}\big)/\Big[\Phi\big(x_{ij}^{\eta}\big)\big\{1-\Phi\big(x_{ij}^{\eta}\big)\big\}\Big] $, where $ x_{ij}^{\eta}=\big(b_j^{\top}\wh{Z}_i^*+\eta_{ij}\wh{\omega}_j^{\top}\wh{\Omega}^{-1}\wh{Z}_i^*-\wh{c}_j\big)/\wh{\tau}_j$. Similar to $\mQ_{32}^b$, we can obtain that $ \mQ_{33}^b\leq 2\big\Vert\wh{Z}_i^*\big\Vert\cdot\big\Vert\wh{\Omega}\big\Vert^{-1}$ $\max_j\Big\vert\wh{\tau}_j^{-1}\phi(x_{ij}^{\eta})/\big[\Phi(x_{ij}^{\eta})\big\{1- \Phi(x_{ij}^{\eta})\big\}\big]\Big\vert\cdot p^{-1}\sum_{j=1}^p\big\Vert\wh{\omega}_j\big\Vert$. Similar to \sc{Step 3.2.1,}\normalfont$\ $one can also verify that $ \big\Vert\wh{Z}_i^*\big\Vert\cdot\big\Vert\wh{\Omega}\big\Vert^{-1}\max_j\Big\vert\wh{\tau}_j^{-1}\phi(x_{ij}^{\eta})/\big[\Phi(x_{ij}^{\eta})\big\{1- \Phi(x_{ij}^{\eta})\big\}\big]\Big\vert=O_p(1) $. Therefore, $\mQ_{33}^b=o_p(1) $ is proved. Then, we have $ \mL_{\wh{B},\tau}^{(i)}\big(\wh{Z}_i\big)\geq \mL_{B,\tau}^{(i)}\big(\wh{Z}_i^*\big)+o_p(1) $. 

\sc{Step 4.}\normalfont$\ $Lastly, we prove that $p^{-1/2}\big\Vert\wh{B}\wh{Z}_i-BZ_i\big\Vert=o_p(1) $. By the proof of \sc{Step 3.2,}\normalfont$\ $we know that $ \wh{b}_j^{\top}\wh{Z}_i=b_j^{\top}\wt{Z}_i^*+\wh{\omega}_j^{\top}\wh{Z}_i$. Direct computation leads to $p^{-1}\big\Vert\wh{B}\wh{Z}_i-BZ_i\big\Vert^2=p^{-1}\sum_{j=1}^p\big(\wh{b}_j^{\top}\wh{Z}_i-b_j^{\top}Z_i\big)^2=p^{-1}\sum_{j=1}^p\big\{b_j^{\top}\big(\wt{Z}_i^*-Z_i\big)+\wh{\omega}_j^{\top}\wh{Z}_i\big\}^2\leq 2\big(\mQ_{41}+\mQ_{42}\big)$, where $\mQ_{41}=\big\Vert\wh{Z}_i\big\Vert^2\cdot p^{-1}\sum_{j=1}^p\big\Vert\wh{\omega}_j\big\Vert^2$ and  $\mQ_{42}=\big\Vert\wt{Z}_i^*-Z_i\big\Vert^2\cdot p^{-1}\sum_{j=1}^p\big\Vert b_j\big\Vert^2$. To prove $ p^{-1/2}\big\Vert\wh{B}\wh{Z}_i-BZ_i\big\Vert =o_p(1) $, it suffices to show that $\mQ_{41}=o_p(1)$ and $\mQ_{42}=o_p(1)$ separately. By \sc{Step 3.1-3.2,}\normalfont$\ $we know that $p^{-1}\sum_{j=1}^p\big\Vert\wh{\omega}_j\big\Vert^2=o_p(1)$ and $\big\Vert\wh{Z}_i\big\Vert^2=O_p(1)$. Consequently,  $\mQ_{41}=o_p(1)$ is proved. By definition, we know that $ \big\Vert b_j\big\Vert^2=1-\tau_{j}^2$ and then $ p^{-1}\sum_{j=1}^p\big\Vert b_j\big\Vert^2\leq1$. Next, we study  $ \big\Vert\wt{Z}_i^*-Z_i\big\Vert\leq \big\Vert \wh{Z}_i^*-Z_i\big\Vert+\big\Vert\wt{Z}_i^*-\wh{Z}_i^*\big\Vert$. By \sc{Step 1-2,}\normalfont$\ $we know that $\big\Vert\wt{Z}_i-Z_i\big\Vert=O_p(1/\sqrt{p})$ and $\sup_{\Vert z\Vert\leq Z_{\max}}\big\vert\mL_{B,\tau}^{(i)}\big(z\big)-\wt{\mL}_{B}^{(i)}\big(z\big)\big\vert=o_p(1)$. Recall that $ \wt{Z}_i=\argmax_{Z_i}\wt{\mL}_{B}^{(i)}\big(Z_i\big) $ and $ \wh{Z}_i^*=\argmax_{Z_i}\mL_{B,\tau}^{(i)}\big(Z_i\big) $. We then can directly obtain that $\big\Vert\wh{Z}_i^*-Z_i\big\Vert\leq \Vert\wh{Z}_i^*-\wt{Z}_i\big\Vert+\Vert\wt{Z}_i-Z_i\big\Vert=o_p(1)$. By \sc{Step 3,}\normalfont$\ $we know that $\big\vert\mL_{B,\tau}^{(i)}\big(\wt{Z}_i^*\big)-\mL_{B,\tau}^{(i)}\big(\wh{Z}_i^*\big)\big\vert=o_p(1)$. In the meanwhile, the log-likelihood function of the probit model is continuous and globally concave in parameters \citep{1984linear,1985advanced}. This implies that $ \wh{Z}_i^*$ is the global maximum point of $ \mL_{B,\tau}^{(i)}\big(Z_i\big)$. Moreover, note that $ \mL_{B,\tau}^{(i)}\big(Z_i\big)$ is a uniform continuous function in $\Vert Z_i\Vert \leq Z_{\max}$. Therefore, we can directly  obtain that $\big\Vert\wt{Z}_i^*-\wh{Z}_i^*\big\Vert=o_p(1) $. Combining the above results, we have $ p^{-1/2}\big\Vert\wh{B}\wh{Z}_i-BZ_i\big\Vert \xrightarrow{p}0$ as $n\to\infty$. This completes the whole proof of Theorem 4.

\section*{Appendix B. Verification Details}

In this Appendix, we show in detail the calculation process of some results given in this paper. 
\subsection*{\textbf{Appendix B.1. Proof of Results in Section 2.1}}

\sc{Part 1.}\normalfont$\ $We need to verify that
$\dot{\ell}\big(\wh{c}_{j_1},\wh{c}_{j_2};\sigma_{j_1j_2}\big)=\partial \ell\big(\wh{c}_{j_1},\wh{c}_{j_2};\sigma_{j_1j_2}\big)/\partial\sigma_{j_1j_2}=\big(1-\sigma_{j_1j_2}^2\big)^{-1/2}\phi\Big\{(\wh{c}_{j_1}-\sigma_{j_1j_2}\wh{c}_{j_2})/(1-\sigma_{j_1j_2}^2)^{1/2}\Big\}\phi\big(\wh{c}_{j_2}\big)$, where $\phi(x)=\big(2\pi\big)^{-1/2}\exp\big(-x^2/2\big)$ stands for the probability distribution function of a standard normal distribution. Recall that $\ell\big(\wh{c}_{j_1},\wh{c}_{j_2};\sigma_{j_1j_2}\big)$ is defined in (2.2). Before computing the first-order partial derivative of $\ell\big(\wh{c}_{j_1},\wh{c}_{j_2};\sigma_{j_1j_2}\big)$ with respect to $\sigma_{j_1j_2}$, we can re-write it as $\int_{\wh{c}_{j_1}}^{\infty}\phi\big(x\big)\int_{\wh{c}_{j_2}(x,\sigma_{j_1j_2})}^{\infty}$  $\phi\big(y\big)dydx$, where $\wh{c}_{j_2}(x,\sigma_{j_1j_2})=(\wh{c}_{j_2}-\sigma_{j_1j_2}x)/\big(1-\sigma_{j_1j_2}^2\big)^{1/2}$ \citep{kotz2004}. Note that $\ell\big(\wh{c}_{j_1},\wh{c}_{j_2};\sigma_{j_1j_2}\big)\leq 1$. Then, we have 
\begin{equation}\label{a.1}
	\dfrac{\partial \ell\big(\wh{c}_{j_1},\wh{c}_{j_2};\sigma_{j_1j_2}\big)}{\partial\sigma_{j_1j_2}}=\int_{\wh{c}_{j_1}}^{\infty}-\phi\big(x\big)\phi\big(\wh{c}_{j_2}(x,\sigma_{j_1j_2})\big)\dfrac{\partial \wh{c}_{j_2}(x,\sigma_{j_1j_2})}{\partial \sigma_{j_1j_2}}dx.
\end{equation}
Direct computation leads to $\phi\big(\wh{c}_{j_2}(x,\sigma_{j_1j_2})\big)=\big(2\pi\big)^{-1/2}\exp\big\{-\big(\wh{c}_{j_2}-\sigma_{j_1j_2}x\big)^2\big/2\big(1-\sigma_{j_1j_2}^2\big)\big\}$ and $\partial \wh{c}_{j_2}(x,\sigma_{j_1j_2})/\partial\sigma_{j_1j_2}=\big(\sigma_{j_1j_2}\wh{c}_{j_2}-x\big)/\big(1-\sigma_{j_1j_2}^2\big)^{3/2}$ \citep{pan2017}. Substituting these results into the right-hand side of (\ref{a.1}), we have
\begin{equation}\label{a.2}
	\dfrac{\partial \ell\big(\wh{c}_{j_1},\wh{c}_{j_2};\sigma_{j_1j_2}\big)}{\partial\sigma_{j_1j_2}}=\dfrac{1}{\sqrt{2\pi}}\phi\big(\wh{c}_{j_2}\big)\int_{\wh{c}_{j_1}}^{\infty}\dfrac{x-\sigma_{j_1j_2}\wh{c}_{j_2}}{\big(1-\sigma_{j_1j_2}^2\big)^{3/2}} \exp\bigg\{-\dfrac{\big(x-\sigma_{j_1j_2}\wh{c}_{j_2}\big)^2}{2\big(1-\sigma_{j_1j_2}^2\big)}\bigg\}dx.
\end{equation}
Let $ t=\big(x-\sigma_{j_1j_2}\wh{c}_{j_2}\big)/\big(1-\sigma_{j_1j_2}^2\big)^{1/2}$, we then have $ t\geq t_{\wh{c}_{j_1}} $, where $t_{\wh{c}_{j_1}}=\big(\wh{c}_{j_1}-\sigma_{j_1j_2}\wh{c}_{j_2}\big)/\big(1-\sigma_{j_1j_2}^2\big)^{1/2}$. Then, the right-hand side of (\ref{a.2}) becomes $\phi\big(\wh{c}_{j_2}\big)\big/\big\{2\pi(1-\sigma_{j_1j_2}^2)\big\}^{1/2}\int_{t_{\wh{c}_{j_1}}}^{\infty}t\exp\big(-t^2/2\big)dt$. It can be directly computed as $\dot{\ell}\big(\wh{c}_{j_1},\wh{c}_{j_2};\sigma_{j_1j_2}\big)=\partial \ell\big(\wh{c}_{j_1},\wh{c}_{j_2};\sigma_{j_1j_2}\big)/\partial\sigma_{j_1j_2}=\big(1-\sigma_{j_1j_2}^2\big)^{-1/2}\phi\Big\{(\wh{c}_{j_1}-\sigma_{j_1j_2}\wh{c}_{j_2})/(1-\sigma_{j_1j_2}^2)^{1/2}\Big\}$  $\phi\big(\wh{c}_{j_2}\big)$. This completes the calculation process of $\dot{\ell}\big(\wh{c}_{j_1},\wh{c}_{j_2};\sigma_{j_1j_2}\big)$.

\sc{Part 2.}\normalfont$\ $We need to show that $\ell\big(\wh{c}_{j_1},\wh{c}_{j_2};\sigma_{j_1j_2}\big)\in \big(\ell\big(\wh{c}_{j_1},\wh{c}_{j_2};-1\big),\ell\big(\wh{c}_{j_1},\wh{c}_{j_2};1\big)\big)$. By \sc{Part 1,}\normalfont$\ $we know that $ \dot{\ell}\big(\wh{c}_{j_1},\wh{c}_{j_2};$ $\sigma_{j_1j_2}\big)>0 $. This implies that $ \ell\big(\wh{c}_{j_1},\wh{c}_{j_2};\sigma_{j_1j_2}\big) $ is a monotonically increasing function of $ \sigma_{j_1j_2}$. Following the idea of \cite{kotz2004}, we can write $ \ell\big(\wh{c}_{j_1},\wh{c}_{j_2};\sigma_{j_1j_2}\big)$ as
\begin{equation*}
	\ell\big(\wh{c}_{j_1},\wh{c}_{j_2};\sigma_{j_1j_2}\big)=\dfrac{1}{2\pi}\int_{\wh{c}_{j_1}}^{\infty}\exp\big(-x^2/2\big)
	\int_{\wh{c}_{j_2}(x,\sigma_{j_1j_2})}^{\infty}\exp\big(-y^2/2\big)dydx,
\end{equation*}
where $ \wh{c}_{j_2}(x,\sigma_{j_1j_2})=(\wh{c}_{j_2}-\sigma_{j_1j_2}x)/(1-\sigma_{j_1j_2}^2)$. Then, it can be verified that $\lim\limits_{\sigma_{j_1j_2}\to\pm1}$  $\ell\big(\wh{c}_{j_1},\wh{c}_{j_2};\sigma_{j_1j_2}\big) =\ell\big(\wh{c}_{j_1},\wh{c}_{j_2};\pm1\big)$. The details are given below.

Recall that $\ell\big(\wh{c}_{j_1},\wh{c}_{j_2};\sigma_{j_1j_2}\big)$ is defined in (2.2). We first consider the values of $\ell\big(\wh{c}_{j_1},\wh{c}_{j_2};\sigma_{j_1j_2}\big)$ when $ \sigma_{j_1j_2}=1$ or -1. If $ \sigma_{j_1j_2}=1$, we know that $ e_{j_1} $ and $ e_{j_2} $ are perfectly positively correlated, i.e., $e_{j_1} = e_{j_2}$. Obviously, $\ell\big(\wh{c}_{j_1},\wh{c}_{j_2};1\big)=1-\Phi\big(\max\big\{\wh{c}_{j_1},\wh{c}_{j_2}\big\}\big)=1-\max\big\{\Phi\big(\wh{c}_{j_1}\big),\Phi\big(\wh{c}_{j_2}\big)\big\}$. If $\sigma_{j_1j_2}=-1$, we know that $ e_{j_1} $ and $ e_{j_2} $ are perfectly negatively correlated, i.e., $e_{j_1} = -e_{j_2}$. Recall that $\ell\big(\wh{c}_{j_1},\wh{c}_{j_2};\sigma_{j_1j_2}\big)= P\big(e_{ij_1}>\wh{c}_{j_1}, e_{ij_2}>\wh{c}_{j_2}\big)$. Then, we can immediately have $\ell\big(\wh{c}_{j_1},\wh{c}_{j_2};-1\big)=1-\Phi\big(\wh{c}_{j_1}\big)-\Phi\big(\wh{c}_{j_2}\big)$ if $\wh{c}_{j_1}+\wh{c}_{j_2}\leq 0$ and $\ell\big(\wh{c}_{j_1},\wh{c}_{j_2};-1\big)=0$ otherwise. 

Next, we prove $\lim\limits_{\sigma_{j_1j_2}\to\pm1}\ell\big(\wh{c}_{j_1},\wh{c}_{j_2};\sigma_{j_1j_2}\big)=\ell\big(\wh{c}_{j_1},\wh{c}_{j_2};\pm1\big)$. Since the proof of $\sigma_{j_1j_2}\to1$ and $\sigma_{j_1j_2}\to-1$ are nearly identical, we shall present
the proof details for $\lim\limits_{\sigma_{j_1j_2}\to1}\ell\big(\wh{c}_{j_1},$  $\wh{c}_{j_2};\sigma_{j_1j_2}\big)$ only. Recall that $\ell\big(\wh{c}_{j_1},\wh{c}_{j_2};\sigma_{j_1j_2}\big)$ is given by
\begin{equation}\label{a.3}
	\ell\big(\wh{c}_{j_1},\wh{c}_{j_2};\sigma_{j_1j_2}\big)=\dfrac{1}{2\pi}\int_{\wh{c}_{j_1}}^{\infty}\exp\big(-x^2/2\big)
	\int_{\wh{c}_{j_2}(x,\sigma_{j_1j_2})}^{\infty}\exp\big(-y^2/2\big)dydx,
\end{equation}
where $ \wh{c}_{j_2}(x,\sigma_{j_1j_2})=(\wh{c}_{j_2}-\sigma_{j_1j_2}x)/(1-\sigma_{j_1j_2}^2)$. Note that $\ell\big(\wh{c}_{j_1},\wh{c}_{j_2};\sigma_{j_1j_2}\big)\leq 1$. Then, $\lim\limits_{\sigma_{j_1j_2}\to1}\ell\big(\wh{c}_{j_1},$  $\wh{c}_{j_2};\sigma_{j_1j_2}\big)$ is determined by $\lim\limits_{\sigma_{j_1j_2}\to1}\wh{c}_{j_2}(x,\sigma_{j_1j_2})$. In other words, we need to focus on the relationship between $ \wh{c}_{j_1} $ and $ \wh{c}_{j_2} $. If $\wh{c}_{j_1}>\wh{c}_{j_2}$, we have $ x\geq \wh{c}_{j_1}>\wh{c}_{j_2}$. Further, we have $\lim\limits_{\sigma_{j_1j_2}\to1}\wh{c}_{j_2}(x,\sigma_{j_1j_2})=-\infty$. Subsequently, $\lim\limits_{\sigma_{j_1j_2}\to1}\ell\big(\wh{c}_{j_1},$  $\wh{c}_{j_2};\sigma_{j_1j_2}\big)=1-\Phi\big(\wh{c}_{j_1}\big)$. If $\wh{c}_{j_1}\leq\wh{c}_{j_2}$, we can divide the right-hand of (\ref{a.3}) into two parts. It means that $\ell\big(\wh{c}_{j_1},\wh{c}_{j_2};\sigma_{j_1j_2}\big) = \ell^{(1)}\big(\wh{c}_{j_1},\wh{c}_{j_2};\sigma_{j_1j_2}\big) + \ell^{(2)}\big(\wh{c}_{j_1},\wh{c}_{j_2};\sigma_{j_1j_2}\big)$, where $  \ell^{(1)}\big(\wh{c}_{j_1},\wh{c}_{j_2};\sigma_{j_1j_2}\big)=\big(2\pi\big)^{1/2}\int_{\wh{c}_{j_1}}^{\wh{c}_{j_2}}\exp\big(-x^2/2\big)
\int_{\wh{c}_{j_2}(x,\sigma_{j_1j_2})}^{\infty}\exp\big(-y^2/2\big)dydx$ and $\ell^{(2)}\big(\wh{c}_{j_1},\wh{c}_{j_2};\sigma_{j_1j_2}\big)=\big(2\pi\big)^{1/2}$  $\int_{\wh{c}_{j_2}}^{\infty}\exp\big(-x^2/2\big)
\int_{\wh{c}_{j_2}(x,\sigma_{j_1j_2})}^{\infty}\exp\big(-y^2/2\big)dydx$. For $\ell^{(1)}\big(\wh{c}_{j_1},\wh{c}_{j_2};\sigma_{j_1j_2}\big)$, we have $ \wh{c}_{j_1}\leq x\leq \wh{c}_{j_2}$, which leads to $\lim\limits_{\sigma_{j_1j_2}\to1}\wh{c}_{j_2}(x,\sigma_{j_1j_2})=\infty$. Then, $\lim\limits_{\sigma_{j_1j_2}\to1}\ell^{(1)}\big(\wh{c}_{j_1},\wh{c}_{j_2};\sigma_{j_1j_2}\big)=0$. For $\ell^{(2)}\big(\wh{c}_{j_1},\wh{c}_{j_2};\sigma_{j_1j_2}\big)$, we have $ x\geq \wh{c}_{j_2}$. Subsequently, we can obtain that $\lim\limits_{\sigma_{j_1j_2}\to1}\ell^{(2)}\big(\wh{c}_{j_1},\wh{c}_{j_2};$  $\sigma_{j_1j_2}\big)=1-\Phi\big(\wh{c}_{j_2}\big)$. Thus, $\lim\limits_{\sigma_{j_1j_2}\to1}\ell\big(\wh{c}_{j_1},$  $\wh{c}_{j_2};\sigma_{j_1j_2}\big)=1-\Phi\big(\wh{c}_{j_2}\big)$. This complete the proof of $\lim\limits_{\sigma_{j_1j_2}\to1}\ell\big(\wh{c}_{j_1},$  $\wh{c}_{j_2};\sigma_{j_1j_2}\big) =\ell\big(\wh{c}_{j_1},$  $\wh{c}_{j_2};1\big) $. Similarly, we can prove $\lim\limits_{\sigma_{j_1j_2}\to-1}\ell\big(\wh{c}_{j_1},$  $\wh{c}_{j_2};\sigma_{j_1j_2}\big)=\ell\big(\wh{c}_{j_1},$  $\wh{c}_{j_2};-1\big)$. Combining the above results, we complete the proof of $\lim\limits_{\sigma_{j_1j_2}\to\pm1}\ell\big(\wh{c}_{j_1},\wh{c}_{j_2};\sigma_{j_1j_2}\big)=\ell\big(\wh{c}_{j_1},\wh{c}_{j_2};\pm1\big)$.

\subsection*{\textbf{Appendix B.2. Proof of Results in Appendix A.4}}

\sc{Step 1.1.}\normalfont$\ $We show that $ p^{1/2}\Big\Vert\dot{\wt{\mL}}^{(i)}_B(Z_i)\Big\Vert=O_p(1)$. Direct computation leads to $p^{1/2}\dot{\wt{\mL}}^{(i)}_B(Z_i)=p^{-1/2}\sum_{j=1}^pb_j\tau_j^{-1}\big\{Y_{ij}-\Phi(x_{ij})\big\}\phi(x_{ij})/\big[\Phi(x_{ij})\big\{1-\Phi(x_{ij})\big\}\big]$ with $x_{ij}=\big(b_j^{\top}Z_i-c_j\big)/\tau_j$. Note that $E\big(Y_{ij}\vert Z_i\big)=\Phi(x_{ij})$, $\var\big(Y_{ij}\vert Z_i\big)=\Phi(x_{ij})\big\{1-\Phi(x_{ij})\big\}$, and $\cov\big(Y_{ij_1},Y_{ij_2}\vert Z_i\big)=0$. We can compute $E\Big\{p^{1/2}\dot{\wt{\mL}}_B^{(i)}(Z_i)\big\vert Z_i\Big\}=0$ and $\var\Big\{p^{1/2}\dot{\wt{\mL}}_B^{(i)}(Z_i)\big\vert Z_i\Big\}=p^{-1}\sum_{j=1}^pb_jb_j^{\top}\tau_j^{-2}\phi^2(x_{ij})/$ $\big[\Phi(x_{ij}\big\{1-\Phi(x_{ij})\big\}\big]$. By Definition, we have $\Vert b_j\Vert \leq 1$. By \sc{Step 2}\normalfont$\ $of Theorem 1 and condition (C5), we have $ \tau_{\min}^2\leq \tau_{j}^2\leq \tau_{\max}^2$ and $\vert x_{ij}\vert\leq x_{\max}$, where $x_{\max}=\tau_{\min}^{-1}\big(Z_{\max}+c_{\max}\big)$. Note that $\vert\phi(x_{ij})\vert\leq \phi(0)$. Consequently, we can obtain that $\big\vert\Phi(x_{ij})\big\{1-\Phi(x_{ij})\big\}\big\vert\geq M_{\Phi}>0$, where $ M_{\Phi}=\Phi(x_{\max})\big\{1-\Phi(x_{\max})\big\}$. Therefore, we have $\Big\Vert p^{-1/2}b_j\tau_j^{-1}\big\{Y_{ij}-\Phi(x_{ij})\big\}\phi(x_{ij})/\big[\Phi(x_{ij})\big\{1-\Phi(x_{ij})\big\}\big]\Big\Vert\leq \phi(0)M_{\Phi}^{-1}\tau_{\min}^{-1}=C_{\ell}$ and $\Big\Vert\var\Big\{p^{1/2}\dot{\wt{\mL}}_B^{(i)}(Z_i)\big\vert Z_i\Big\}\Big\Vert\leq \phi^2(0)M_{\Phi}^{-1}\tau_{\min}^{-2}=C_{\ell}\phi(0)$ as long as $p>1$. Let $U=p^{1/2}\dot{\wt{\mL}}_B^{(i)}(Z_i)$. By \cite{shao2003}, we know that $\var\big(U\big)=E\big\{\var\big(U\vert Z_i\big)\big\}+\var\big\{E\big(U\vert Z_i\big)\big\}$. Recall that $E\big(U\vert Z_i\big)=0$ and $\big\Vert\var\big(U\vert Z_i\big)\big\Vert\leq C_{\ell}\phi(0)$. We then have $\big\Vert\var\big(U\big)\big\Vert\leq  C_{\ell}\phi(0)$. Therefore, we can obtain that $E\big(U^{\top}U\big)=\tr\Big\{\var\big(U\big)\Big\}\leq d\big\Vert\var\big(U\big)\big\Vert\leq  dC_{\ell}\phi(0)$. Thus, for any sufficiently large $M>0$, the Markov's Inequality \citep{1892functions,1985markoff} can be applied as
\begin{equation}\label{c4}
	P\bigg\{\Big\Vert p^{1/2}\dot{\wt{\mL}}^{(i)}_B(Z_i)\Big\Vert^2>M\bigg\}\leq E\Big(U^{\top}U\Big)/M\leq dC_{\ell}\phi(0)/M.
\end{equation}
Therefore, we have $ p^{1/2}\Big\Vert\dot{\wt{\mL}}^{(i)}_B(Z_i)\Big\Vert=O_p(1)$.

\sc{Step 2.1.}\normalfont$\ $We start with $ \sup_{\Vert z\Vert\leq Z_{\max}}\max_j\vert\mQ_{21}\vert$. Recall that $\mQ_{21}=-p^{-1}\sum_{j=1}^pI\big(\wh{\tau}_j\leq\tau\big)\ell_{B,\tau}^{(i,j)}\big(z,\zeta_j\big)$, $\ell_{B,\tau}^{(i,j)}\big(z,\zeta_j\big)=Y_{ij}\log\Phi(x_{ij})+\big(1-Y_{ij}\big)\log\big\{1-\Phi(x_{ij})\big\}$ and $x_{ij}=\big(b_j^{\top}z-c_j\big)/\tau_j$. By \sc{Step 1}\normalfont$\ $of Appendix B.4, we can obtain that $\vert x_{ij}\vert\leq x_{\max}$ and then $\vert\ell_{B,\tau}^{(i,j)}\big(z,\zeta_j\big)\vert\leq 2\big\vert\log\Phi\big(x_{\max}\big)\big\vert$. We further have $\sup_{\Vert z\Vert\leq Z_{\max}}\max_j\vert\mQ_{21}\vert\leq  2\big\vert\log\Phi\big(x_{\max}\big)\big\vert$ $ p^{-1} \sum_{j=1}^pI\big(\wh{\tau}_j\leq\tau\big)$. By Theorem 3, we know that $ p^{-1}\sum_{j=1}^p\big\vert\wh{\tau}_j^2-\tau_j^2\big\vert=o_p(1)$ as $n\to\infty$. In the meanwhile, we have $p^{-1}\sum_{j=1}^p\big\vert\wh{\tau}_j^2-\tau_j^2\big\vert\geq p^{-1}\sum_{j=1}^pI\big(\wh{\tau}_j\leq\tau\big)\big\vert\wh{\tau}_j^2-\tau_j^2\big\vert\geq\big(\tau_{\min}^2-\tau^2\big)p^{-1}\sum_{j=1}^pI\big(\wh{\tau}_j\leq\tau\big)$. Consequently, we know that $ p^{-1}\sum_{j=1}^pI\big(\wh{\tau}_j\leq\tau\big)=o_p(1)$ as $n\to\infty$. Therefore, we have $ \sup_{\Vert z\Vert\leq Z_{\max}}\max_j\vert\mQ_{21}\vert =o_p(1)$ as $n\to\infty$. 

\sc{Step 2.2.}\normalfont$\ $Next, we prove $\sup_{\Vert z\Vert\leq Z_{\max}}\max_j\vert\mQ_{22}\vert=o_p(1)$ as $n\to\infty$. Recall that $\mQ_{22}= p^{-1}\sum_{j=1}^pI\big(\wh{\tau}_j>\tau\big)\big\{\dot{\ell}_{B,\tau}^{(i,j)}\big(z,\wt{\zeta}_j\big)\big\}^{\top}$ $\big(\wh{\zeta}_j-\zeta_j\big)$, $\zeta_j=\big(c_j,\tau_j^2\big)^{\top}\in\mR^2$ and $ \wt{\zeta}_j=\alpha_j\wh{\zeta}_j+(1-\alpha_j)\zeta_j$. We then have $\sup_{\Vert z\Vert\leq Z_{\max}}\max_j\vert\mQ_{22}\vert\leq \sup_{\Vert z\Vert\leq Z_{\max}}\max_j\big\Vert \dot{\ell}_{B,\tau}^{(i,j)}\big(z,\wt{\zeta}_j\big)\big\Vert\cdot\Big\{p^{-1}\sum_{j=1}^pI\big(\wh{\tau}_j>\tau\big)\big\Vert\wh{\zeta}_j-\zeta_j\big\Vert\Big\}$. By Section 2 in this paper, we know that $\max_j\big\vert\wh{c}_j-c_j\big\vert =o_p(1)$ and $p^{-1}\sum_{j=1}^p\big\vert\wh{\tau}_j^2-\tau_j^2\big\vert=o_p(1)$ as $n\to\infty$. Thus, we can obtain that $ p^{-1}\sum_{j=1}^pI\big(\wh{\tau}_j>\tau\big)\big\Vert\wh{\zeta}_j-\zeta_j\big\Vert\leq p^{-1}\sum_{j=1}^p\big\vert\wh{c}_j-c_j\big\vert+p^{-1}\sum_{j=1}^p\big\vert\wh{\tau}_j^2-\tau_j^2\big\vert\xrightarrow{p}0$ as $n\to\infty$. Next, we focus on $\sup_{\Vert z\Vert\leq Z_{\max}}\max_j\big\Vert \dot{\ell}_{B.,\tau}^{(i,j)}\big(z,\wt{\zeta}_j\big)\big\Vert$. Recall that $ \ell_{B,\tau}^{(i,j)}\big(z,\zeta_j\big)=\Big(Y_{ij}\log$ $\Phi\big\{\big(b_j^{\top}z-c_j\big)/\tau_j\big\}+\big(1-Y_{ij}\big)\log\Big[1-\Phi\big\{\big(b_j^{\top}z-c_j\big)/\tau_j\big\}\Big]\Big)$ and $0<\tau<\tau_{\min}$. Direct computation leads to $ \dot{\ell}_{B,\tau}^{(i,j)}\big(z,\zeta_j\big)=\big\{Y_{ij}-\Phi(x_{ij})\big\}\dot{\Phi}(x_{ij})/\big[\Phi(x_{ij})\big\{1-\Phi(x_{ij})\big\}\big]$, where $ \dot{\Phi}(x_{ij})=\partial \Phi(x_{ij})/\partial\zeta_j\in\mR^2$ and $x_{ij}=\big(b_j^{\top}z-c_j\big)/\tau_j$. We then have $\wt{x}_{ij}=\big(b_j^{\top}z-\wt{c}_j\big)/\wt{\tau}_j$. By definition, we can upper bound $\vert\wt{x}_{ij}\vert$ as $\vert\wt{x}_{ij}\vert\leq \wt{x}_{\max} $ with $\wt{x}_{\max}=\tau^{-1}\big(Z_{\max}+c_{\max}\big)$. Note that $ Y_{ij}\in\{0,1\}$ and $0\leq \Phi(\cdot)\leq 1$. Thus, we have $ \big\Vert \dot{\ell}_{B.,\tau}^{(i,j)}\big(z,\wt{\zeta}_j\big)\big\Vert\leq 2\Phi^{-1}\big(\wt{x}_{\max}\big)\Phi^{-1}\big(-\wt{x}_{\max}\big)\big\Vert\dot{\Phi}\big(\wt{x}_{ij}\big)\big\Vert$. Write $ \dot{\Phi}\big(\wt{x}_{ij}\big)=\big(\dot{\Phi}^{c}(\wt{x}_{ij}),\dot{\Phi}^{\tau_j^2}(\wt{x}_{ij})\big)^{\top}$, where $ \dot{\Phi}^{c}(\wt{x}_{ij})=-\wt{\tau}_j^{-1}\phi(\wt{x}_{ij})$ and $\dot{\Phi}^{\tau_j^2}(\wt{x}_{ij})=-2^{-1}\wt{x}_{ij}\wt{\tau}_j^{-2}\phi(\wt{x}_{ij})$. We then have $\big\Vert\dot{\Phi}\big(\wt{x}_{ij}\big)\big\Vert\leq \big(1+2^{-1}\wt{x}_{\max}$ $\tau^{-1}\big)\tau^{-1}\phi\big(0\big)$. Therefore, we can obtain that $ \big\Vert \dot{\ell}_{B.,\tau}^{(i,j)}\big(z,\wt{\zeta}_j\big)\big\Vert\leq \ell_{\max}$, where $\ell_{\max}=2\Phi^{-1}\big(\wt{x}_{\max}\big)\Phi^{-1}\big(-\wt{x}_{\max}\big)\big(1+2^{-1}\wt{x}_{\max}\tau^{-1}\big)\tau^{-1}\phi\big(0\big)$. This completes the proof of $\sup_{\Vert z\Vert\leq Z_{\max}}$ $\max_j\vert\mQ_{22}\vert=o_p(1)$ as $n\to\infty$.

\sc{Step 3.2.1.}\normalfont$\ $ Recall that  $\wh{\Omega}=\big(B^{\top}B\big)^{-1}B^{\top}\wh{B}=\big(B^{\top}B/p\big)^{-1}B^{\top}\wh{\Sigma}_d\wh{\Lambda}_d^{1/2}/p$. We then have $\big\Vert\wh{\Omega}\big\Vert\leq \big\Vert\Sigma_B\big\Vert^{-1}\cdot\big\Vert B/\sqrt{p}\big\Vert\cdot\big\Vert\wh{\Sigma}_d\wh{\Lambda}_d^{1/2}/\sqrt{p}\big\Vert$. Note that $\big\Vert\Sigma_B\big\Vert^{-1}=\lambda_{\min}^{-1}\big(\Sigma_B\big)=\lambda_d^{-1}\big(\Sigma_B\big) $ and $ \big\Vert B/\sqrt{p}\big\Vert=\lambda_{\max}^{1/2}\big(BB^{\top}/p\big)=\lambda_1\big(\Sigma_B\big)$. In the meanwhile, we can obtain that  $\big\Vert\wh{\Sigma}_d\wh{\Lambda}_d^{1/2}/\sqrt{p}\big\Vert^2=\lambda_{\max}\big(\wh{\Sigma}_d\wh{\Lambda}_d\wh{\Sigma}_d^{\top}/p\big)\leq \lambda_{\max}\big(\wh{\Sigma}/p\big)=\lambda_{\max}\big(\Sigma/p\big)+\big\Vert\wh{\Sigma}-\Sigma\big\Vert/p \leq \lambda_1\big(\Sigma_B\big)+\tau_{\max}^2/p+o_p(1)$. Thus, we can immediately have $ \big\Vert\wh{\Omega}\big\Vert\leq \Omega_{\max}+o_p(1)$, where $\Omega_{\max}= \lambda_d^{-1}\big(\Sigma_B\big)\lambda_1^{1/2}\big(\Sigma_B\big)$ $\big\{\lambda_1\big(\Sigma_B\big)+\tau_{\max}^2\big\}^{1/2}$. Recall that $\wt{Z}_i^*=\wh{\Omega}\wh{Z}_i$. By condition (C5) and section 2.4, we can obtain that $ \big\Vert Z_i\big\Vert\leq Z_{\max}$ and $\big\Vert \wh{Z}_i\big\Vert=O_p(1)$. Consequently, we know that $ \big\Vert\wt{Z}_i^*\big\Vert=O_p(1)$. Recall that $ x_{ij}^{\alpha}=\big(b_j^{\top}\wt{Z}_i^*+\alpha_{ij}\wh{\omega}_j^{\top}\wh{Z}_i-\wh{c}_j\big)/\wh{\tau}_j$ and $ \wh{\omega}_j^{\top}=\big(\delta_j^H\big)^{\top}\wh{B}$. We then have $\big\vert x_{ij}^{\alpha} \big\vert\leq \big(\big\Vert\wt{Z}_i^*\big\Vert+\big\Vert\wh{\omega}_j\big\Vert\cdot\big\Vert\wh{Z}_i\big\Vert+c_{\max}\big)/\tau=O_p(1)$ provided $\wh{\tau}_j>\tau$. As a result, $\Big\vert\phi(x_{ij}^{\alpha})/\big[\Phi(x_{ij}^{\alpha})\big\{1- \Phi(x_{ij}^{\alpha})\big\}\big]\Big\vert=O_p(1) $. Therefore, we have proved that
$\big\Vert\wh{Z}_i\big\Vert\max_j\Big\vert I\big(\wh{\tau}_j>\tau\big)\wh{\tau}_j^{-1}\phi(x_{ij}^{\alpha})/$ $\big[\Phi(x_{ij}^{\alpha})\big\{1- \Phi(x_{ij}^{\alpha})\big\}\big]\Big\vert=O_p(1)$.

\bibliographystyle{myjmva}
\bibliography{trial}

\end{document}